\documentclass[aps,prd,12pt,nofootinbib,showpacs,floatfix]{revtex4}
\usepackage[dvips]{graphicx}

\begin{document}
\newcommand{\tr}{\mbox{tr}\,}
\newcommand{\Dslash}{{\mathchoice
    {\Dslsh \displaystyle}%
    {\Dslsh \textstyle}%
    {\Dslsh \scriptstyle}%
    {\Dslsh \scriptscriptstyle}}}
\newcommand{\Dslsh}[1]{\ooalign{\(\hfill#1/\hfill\)\crcr\(#1D\)}}
\newcommand{\leftvec}[1]{\vect \leftarrow #1 \,}
\newcommand{\rightvec}[1]{\vect \rightarrow #1 \:}
\renewcommand{\vec}[1]{\vect \rightarrow #1 \:}
\newcommand{\vect}[3]{{\mathchoice
    {\vecto \displaystyle \scriptstyle #1 #2 #3}%
    {\vecto \textstyle \scriptstyle #1 #2 #3}%
    {\vecto \scriptstyle \scriptscriptstyle #1 #2 #3}%
    {\vecto \scriptscriptstyle \scriptscriptstyle #1 #2 #3}}}
\newcommand{\vecto}[5]{\!\stackrel{{}_{{}_{#5#2#3}}}{#1#4}\!}
\newcommand{\vdot}{\!\cdot\!}

\bibliographystyle{apsrev}

\preprint{CU-TP-1146}

\title{Non-perturbatively Determined Relativistic Heavy Quark Action}
\author{Huey-Wen Lin}
\email{hwlin@phys.columbia.edu} \affiliation{Physics
Department, Columbia University, New York, NY 10027, USA}
\author{Norman Christ}
\email{nhc@phys.columbia.edu} \affiliation{Physics Department,
Columbia University, New York, NY 10027, USA}
\date{August 7, 2006}

\begin{abstract}
We present a method to non-perturbatively determine the parameters
of the on-shell, $O(a)$-improved relativistic heavy quark action.
These three parameters, $m_0$, $\zeta$, and $c_B=c_E$ are obtained by
matching finite-volume, heavy-heavy and heavy-light meson masses to
the exact relativistic spectrum through a finite-volume, step-scaling
recursion procedure.  We demonstrate that accuracy on the level
of a few percent can be achieved by carrying out this matching
on a pair of lattices with equal physical spatial volumes but quite
different lattice spacings.  A fine lattice with inverse lattice
spacing $1/a=5.4$ GeV and $24^3 \times 48$ sites and a coarse,
$1/a=3.6$ GeV, $16^3 \times 32$ lattice are used together with
a heavy quark mass $m$ approximately that of the charm quark.
This approach is unable to determine the initially expected,
four heavy-quark parameters: $m_0$, $\zeta$, $c_B$ and $c_E$.
This apparent non-uniqueness of these four parameters motivated the
analytic result, presented in a companion paper, that this set is
redundant and that the restriction $c_E=c_B$ is permitted through
order $a|\vec p|$ and to all orders in $m a$ where $\vec p$ is the
heavy quark's spatial momenta.
\end{abstract}

\maketitle

\section{Introduction}

The study of flavor physics and CP violation plays a central role
in particle physics.  In particular, many of the parameters of
the Standard Model can be constrained by measurements of the
properties of hadrons containing heavy quarks.  However, to do
this one needs theoretical determination of strong-interaction
masses and matrix elements to connect the experimental measurements
with those fundamental quantities.  Lattice quantum chromodynamics
(QCD) provides a first-principles method for the computation of
these hadronic masses and matrix elements.  However, lattice
calculations with heavy quarks present special difficulties since
in full QCD calculations, which properly include the effects of
dynamical quarks, it is impractical to work with lattice spacings
sufficiently small that errors on the order of $ma$ can be
controlled.  These problems are addressed by using a number of
improved heavy quark actions designed to control or avoid these
potentially important finite lattice spacing errors.  The results
of recent calculations of basic parameters of the Standard Model
can be found in the lattice heavy quark reviews of
Refs.~\cite{Ryan:2001ej,Yamada:2002wh,Kronfeld:2003sd,
Wingate:2004xa,Okamoto:2005zg}.

A variety of fermion actions have been used in lattice calculations
involving heavy quarks~\cite{Mannel:1991mc,Thacker:1990bm,
Lepage:1992tx,Hashimoto:1995in,Sloan:1997fc,Foley:2002qv}.  These
include heavy quark effective theory (HQET) (for which the static
approximation is the leading term) and non-relativistic QCD (NRQCD).
These methods have significant limitations: NRQCD has no continuum
limit, and although HQET has a continuum limit, it cannot be applied
to quarkonia.  While systems involving bottom quarks may permit a
successful expansion in inverse powers of $m$, this is likely not
true for systems including a charm quark.

A third approach, the one adopted here, is the Fermilab or
relativistic heavy quark (RHQ) method~\cite{El-Khadra:1997mp,
Aoki:2001ra} in which extra axis-interchange asymmetric terms
are added to the usual relativistic action.  As is discussed
below, this action can accurately describe heavy quark systems
provided the improvement coefficients it contains are properly
adjusted.  As the heavy quark mass decreases, this action goes
over smoothly to the order $a$ improved fermion action of
Sheikholeslami and Wohlert (SW)~\cite{Sheikholeslami:1985ij}.
Thus, it seems appropriate to refer to this as the relativistic
heavy quark method since it retains the relativistic form of
the Wilson fermion action (with the exception that lattice axis
interchange symmetry is broken) and approaches the standard
relativistic action as $ma$ becomes small.

As is discussed in the original papers~\cite{El-Khadra:1997mp,
Aoki:2001ra} and considered in detail in our companion
paper~\cite{3parAction}, this approach builds upon the
original work of Sheikholeslami and Wohlert, extending it to
to the case of a possibly very heavy quark with mass $m \ge 1/a$
but restricted to a reference system in which these quarks are
nearly at rest.  Such a situation can be described by a Symanzik
effective Lagrangian which contains terms of higher dimension
than four which explicitly reproduce the finite lattice spacing
errors implied by the lattice Lagrangian.

In this RHQ approach, the continuum effective Lagrangian is imagined
to reproduce errors of first order in $a|\vec p|$ and all orders
in $m a$ or $p_0 a$ where $(\vec p, p^0)$ is the heavy quark
four-momentum.  Such an effective action will contain many terms,
including those with arbitrarily large powers of the combination
$a D^0$ where the gauge-covariant time derivative $D^0$ will
introduce a factor of $p^0$ and so cannot be neglected.  As
described in Ref.~\cite{El-Khadra:1997mp} and discussed in detail
in our companion paper, this Lagrangian can be greatly simplified
by performing field transformations within the path integral for
the effective theory.  These field transformation do not change
the particle masses predicted by the theory and, as is discussed
in Ref.~\cite{3parAction}, they effect on-shell spinor
Green's functions only by a simple, Lorentz non-covariant,
$4 \times 4$ spinor rotation.  (The use of the equations of motion
in Ref.~\cite{Aoki:2001ra} is equivalent to the above field
transformation approach.)

As is shown in our companion paper~\cite{3parAction}, after
this simplification, the resulting effective Lagrangian contains
only three parameters: the quark mass $m^c$, an asymmetry parameter
$\zeta^c$ describing the ratio of the coefficients of the spatial
and temporal derivative and a generalization of the Sheikholeslami
and Wohlert, $c_{SW}$ to the case of non-zero mass which we refer
to as $c_P^c$.  Here the superscript $c$ indicates that these are
the coefficients that appear in the continuum effective Lagrangian.
If these three, mass-dependent parameters can be tuned to physical
continuum values ($m^c$, 1 and 0 respectively) by the proper choice
of mass-dependent coefficients in the lattice action, then the
hadronic masses computed in the resulting theory will contain errors
no larger than $(\vec p a)^2$.

In addition, a new parameter $\delta$ multiplying a non-covariant
$\vec \gamma \cdot \vec p$ in the $4 \times 4$ spinor matrix mentioned
above will be needed to realize truly covariant on-shell Greens
functions.  Here $\delta$ will depend on the (usually composite)
fermion operator being used even for a fixed action.  As is discussed
below and in detail in Ref.~\cite{3parAction}, this is one fewer
parameter than found in the previous work of the Fermilab and Tsukuba
groups.

Thus, in our calculation we use the relativistic heavy quark lattice
action:
\begin{eqnarray}\label{eq:RHQ}
S_{\rm lat} &=& \sum_{n',n} \overline{\psi}_{n'} \Bigl(
         \gamma_0 D_0 + \zeta \vec{\gamma} \cdot \vec{D}
         + m_0 - \frac{1}{2} r_t D_0^2 - \frac{1}{2} r_s \vec{D}^2 \nonumber \\
      && + \sum_{i,j} \frac{i}{4} c_B \sigma_{ij}F_{ij}
         + \sum_{i}  \frac{i}{2} c_E \sigma_{0i}F_{0i}
        \Bigr)_{n',n}\psi_n,
\end{eqnarray}
where
\begin{eqnarray}
D_{\mu}\psi(x) & = & \frac{1}{2}\left[U_{\mu}(x)\psi(x +\hat{\mu})
-
U_{\mu}^{\dagger}(x-\hat{\mu})\psi(x-\hat{\mu})\right]\\
D_{\mu}^2\psi(x)& = &\left[U_{\mu}(x)\psi(x+\hat{\mu})
+U_{\mu}^{\dagger}(x-\hat{\mu})\psi(x-\hat{\mu})-2\psi(x)\right]\\
F_{\mu\nu}\psi(x)&=& \frac{1}{8} \sum_{s,s^\prime=\pm
1}\!\!\!\!
ss'\left[U_{s\mu}(x)U_{s^\prime \nu}(x+s\hat\mu)\right.
         U_{-s\mu}(x+s\hat\mu+s^\prime \hat\nu)\nonumber\\
&& \quad\quad\quad\left.\times U_{-s^\prime\nu}(x+s^\prime \hat\nu) -\mathrm{h.c.}\right]\psi(x)
\end{eqnarray}
and the $\gamma$ matrices satisfy: $\gamma_\mu = \gamma_\mu^\dagger$,
$\{\gamma_\mu, \gamma_\nu\} = 2\delta_{\mu\nu}$ and $ \sigma_{\mu\nu} =
\frac{i}{2}[\gamma^\mu,\gamma^\nu]$.
Written in this standard form, there are six possible parameters
that can be adjusted to improve the resulting long-distance theory,
three more than are needed.  We begin by making the choice
$r_s=\zeta$ and $r_t=1$.  This leaves four parameters whose effects we can
study.  In the following we will investigate the non-perturbative
effects of these four parameters, $m_0$, $\zeta$, $c_B$ and $c_E$.
However, when determining an improved RHQ action non-perturbatively,
we will impose the further restriction $c_B = c_E$, making the
improved action uniquely defined at our order of approximation.

The different coefficient choices of improved lattice action
by the Fermilab and Tsukuba groups yield two distinct sets of
coefficients for the action.  These are summarized in
Table~\ref{tab:RHQComp}. The coefficients in each approach
have been calculated by applying lattice perturbation theory
at the $O(a)$-improved, one-loop level to the quark propagator and
quark-quark scattering amplitude~\cite{Aoki:2003dg,Nobes:2005dz}.

In this paper, we will propose and demonstrate a non-perturbative
method for determining these coefficients based on a step-scaling
approach, which eliminates all errors of $O\left(g^{2n}\right)$.
Step scaling has been used in the past to connect the lattice
spacing accessible in large-volume simulations with a lattice scale
sufficiently small that perturbation theory becomes
accurate~\cite{Luscher:1991wu}.  Non-perturbative matching conditions
are imposed to connect the original calculation with one performed
at a smaller lattice spacing $a^\prime_1 = a \epsilon$.  Iterative
matching of this sort with $n$ steps then connects the theory of
interest and a target lattice theory defined with lattice spacing
$a^\prime_n = a \epsilon^n$.   This may require a number of steps
$n$ which is not too large since while the coupling constant
decreases only logarithmically with the energy scale, that energy
scale increases exponentially with $n$.  For example, if a final
comparison with order $g^2$ perturbation theory is used, we expect
an error of order $(g^\prime)^4 \sim \ln(a^\prime_n)^{-2} \sim n^{-2}$
where $g^\prime$ is the bare coupling for the finest lattice.

The situation for heavy quarks is even more favorable.  Here the
target, comparison theory does not need to have such small lattice
spacing that perturbation theory is accurate.  In fact, this theory
can be treated non-perturbatively provided the final lattice
spacing $a^\prime_n$ is sufficiently small that simulations with an
ordinary $O(a)$-improved relativistic action will give accurate
results~\cite{Heitger:2003nj}.  This implies that the size of the
error will be of order $(m a^\prime)^2 \sim n^{-2}$ or
$n \sim (\rm error)^{-1/2}$.  In the work reported here we will
match the step-scaled heavy quark theory with an $O(a)$-accurate
lattice calculation performed using domain-wall fermions.

A critical question in developing such a step-scaling approach
is to decide upon the actual quantities that will be ``matched''
when comparing two theories defined with different lattice spacings
but which are intended to be physically equivalent.  Among the
quantities which might be matched are the Schr\"odinger
functional~\cite{Luscher:1991wu}, off-shell Green's functions
defined in the RI/MOM scheme~\cite{Martinelli:1995ty} or physical
masses and matrix elements at finite volume.

Our first approach to this topic was to investigate the off-shell
RI/MOM scheme since this method had worked well in earlier light-quark
calculations, see {\it e.g.} Refs.~\cite{Blum:2001sr,Blum:2001xb} and
also permits a direct comparison with quantities defined in perturbation
theory.  We were able to define RI/MOM kinematics which lay within
the regime of validity of the effective heavy quark theory described
above and to carry out a tree-level calculation of the amplitudes of
interest~\cite{Lin:2004ht,Lin:2005zd}.  However, the increased number
of parameters needed in the effective theory to describe off-shell
amplitudes, the need to work with gauge-non-invariant quantities
and the difficulty of computing ``disconnected'' gluon correlation
functions ultimately made this approach appear impractical.

In this paper, we adopt the third method mentioned above and
determine the coefficients in the heavy quark effective action
appearing in our step-scaling procedure by requiring that the
physical, momentum-dependent mass spectrum of two physically
equivalent theories agree when compared on the same physical
volume.  Since the step-scaling approach requires physically
small volumes be studied, these spectra will be significantly
distorted by the effects of finite volume and it is important
that these effects be the same in each of the theories being
compared---thus the need to compare on identical physical
volumes.  By comparing more physical, finite-volume quantities
(as many as seven) than there are parameters to adjust (three),
we also have an over-all consistency test of the method.
Finally, as described above, at the smallest lattice spacing,
we compute the quantities being compared using a standard
domain wall fermion action which is has no order $ma$ errors
and accurate chiral symmetry.  We assume that at this smallest
lattice spacing the explicit errors of order $(ma)^2$, present
in the domain wall fermion calculation, are sufficiently small
to be neglected.  Preliminary results using this method were
published in Ref.~\cite{Lin:2005ze}.

The structure of this paper is as follows. We introduce our
on-shell approach to determine the coefficients in the
relativistic heavy quark action via step-scaling both in the
quenched approximation and for full QCD in Sec.~\ref{sec:Strategy}.
In this paper we will work in the quenched approximation and
explicitly carry out the first stage of matching between a
fine and a coarse lattice in order to determine the feasibility
of this approach and the accuracy that can be achieved.
Specifically the ``fine'' lattice uses $1/a=5.4$ GeV and the
``coarse'' lattice $1/a=3.6$ GeV.  (A second matching step,
evaluating the first coarse lattice action on a larger physical
volume and matching to an even larger lattice spacing, $1/a=2.4$ GeV,
is now underway.)  Section~\ref{Sec:simulations} lists the
parameters used in this calculation, describes our determination
of the lattice spacing and discusses our method for obtaining
the physical heavy-heavy and heavy-light spectrum.

The problem of determining the parameters to be used in the
coarse-lattice effective action which will reproduce the fine
lattice mass spectrum is studied in Sec.~\ref{Sec:Analysis}
and the dependence of this spectrum on the four parameters
$m_0$, $\zeta$, $c_B$ and $c_E$ presented.  We are unable to
determine these four parameters with any reasonably precision,
a conclusion we now understand since only three parameters
are required to determine the mass spectrum to order
$a|\vec p|$ and all orders in $(ma)^n$~\cite{3parAction}.
We then restrict the parameter space to $c_B=c_E$, as is
justified theoretically, and find that the resulting three
parameters can now be determined quite accurately.  In
Sec.~\ref{Sec:comp} we compare our result with both perturbative
and non-perturbative determinations of the quark mass and the
one-loop lattice perturbation calculation of the lattice
parameters $\zeta$, $c_B$ and $c_E$ performed by
Nobes~\cite{Nobes:thesis}.  Section.~\ref{SecFuture} presents a
summary and outlook for this approach.

\section{Strategy}
\label{sec:Strategy}

\begin{figure}[hbt]
\includegraphics[width=\columnwidth]{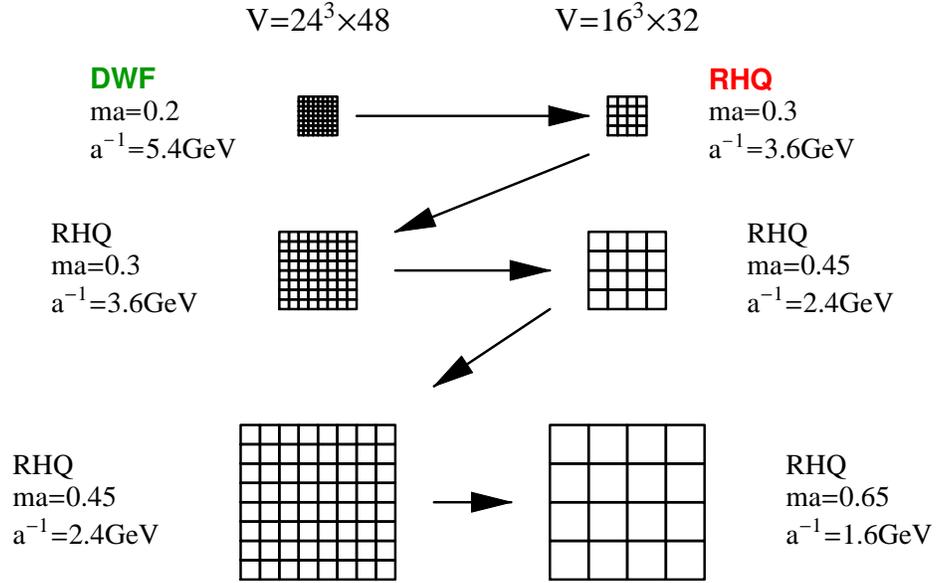}
\caption{The sequence of lattice sizes and lattice spacings used
to determine the coarse-lattice, heavy quark parameters through
a step-scaling technique beginning with a comparison with an
$O(a)$-improved light quark calculation.  The matching between
the top two lattice spacings is the calculation described in this
paper.}
\label{fig:step_scaling} \vskip -3mm
\end{figure}

We propose to determine the three coefficients $m_0$, $\zeta$ and
$c_P \equiv c_B=c_E$ in the RHQ lattice action of Eq.~\ref{eq:RHQ} by
carrying out a series of matching steps.  We begin with a sufficiently
fine lattice spacing that no heavy quark improvements are needed
($ma \ll 1$) and a conventional light-quark calculation will give
accurate results.  We then carry out a series of calculations using
the RHQ lattice action of Eq.~\ref{eq:RHQ} on lattices with increasingly
large lattice spacing and increasingly large physical volume.  When we
increase the lattice spacing at fixed physical volume, we perform
calculations at both lattice spacings on identical physical volumes and
require that the resulting finite-volume, heavy-heavy and heavy-light
energy spectrum agree when these particles are at rest or have small
spatial momenta.  When we increase the lattice volume at fixed lattice
spacing, we simply use the parameters, previously determined, in a
calculation now on the larger volume.  The first same-physical-volume
matching of the energy spectrum is done between the heavy quark theory and
a conventional fine-lattice calculation done with domain wall fermions.
An example of this finite volume, step-scaling recursion is shown in
Fig.~\ref{fig:step_scaling}.

Specifically, we will calculate the pseudo-scalar~(PS), vector~(V),
scalar~(S), and axial-vector~(AV) meson masses in the heavy-heavy~($hh$)
system and pseudo-scalar and vector masses for the heavy-light~($hl$)
system.  We will work with the following mass combinations:
\begin{itemize}
\item Spin-average:  $m^{hh}_{\rm sa}=\frac{1}{4}\left(m^{hh}_{\rm PS} + 3
m^{hh}_{\rm V}\right)$,
$m^{hl}_{\rm sa}=\frac{1}{4}\left(m^{hl}_{\rm PS}+ 3 m^{hl}_{\rm V}\right)$
\item Hyperfine splitting: $m^{hh}_{\rm hs}=m^{hh}_{\rm V} - m^{hh}_{\rm PS}
$, $m^{hl}_{\rm hs}= m^{hl}_{\rm V}-m^{hl}_{\rm PS}$
\item Spin-orbit average and splitting: $m^{hh}_{\rm soa}
=\frac{1}{4}\left(m^{hh}_{\rm S} + 3 m^{hh}_{\rm AV}\right)$,
$m^{hh}_{\rm sos}=m^{hh}_{\rm AV} - m^{hh}_{\rm S} $
\item Mass ratio: $m_1/m_2$ where $E^2 = m_1^2 + \frac{m_1}{m_2} p^2$,
with $m_1$ the rest mass and $m_2$ the kinetic mass.
\end{itemize}
By examining these seven quantities we should be able to determine
the three parameters $m_0$, $\zeta$ and $c_P$ and also check the
size of the scaling violations.

The first step in this program calculates these seven quantities
using the domain wall fermion action on the fine, $24^3 \times 64$
lattice with $1/a=5.4$ GeV (I).  Next, these seven quantities are
computed a second time using a coarse, $16^3 \times 32$ lattice with
$1/a=3.6$ GeV and, therefore, the same physical volume.  This is
calculation II.  The three heavy-quark parameters entering this
coarse-lattice calculation must be adjusted so that these seven
quantities agree between calculations I and II.  It is these
calculations that are carried out in this paper using the parameters
given in Table~\ref{tab:parametersA}.

Third, we expand the volume of calculation II to $24^3\times 48$,
while keeping all other parameters fixed.  The results of this third,
expanded volume calculation (III) can then be matched with a fourth
calculation which has a lattice spacing larger by a factor of 3/2 (IV).  The
simulation parameters for this second matching step are given in
Table~\ref{tab:parametersB}.  By repeating this pattern, we can extend
the calculation to quenched lattices with the desired volume where
serious, infinite-volume, charm physics may be studied.

In this paper, we demonstrate only the matching between calculations
I and II.  The leading heavy quark discretization error in calculation I
is $(ma)^2 \approx 4\%$, making it the dominant systematic error on the
fine lattice result.  Of course, this error can be reduced in future
calculations by choosing a fine lattice that has an even smaller lattice
spacing and correspondingly smaller physical volume.  Without improvement
beyond the usual Sheikholeslami and Wohlert term, the leading heavy
quark discretization error on the coarse lattice is expected to be
$(ma)^2 \approx 10\%$.  However, once we introduce the improved lattice
action of Eq.~\ref{eq:RHQ} and properly tune the coefficients, we
should be able to reduce the error to $(a|\vec p|)^2 \approx 1\%$.

As will be demonstrated in the remainder of this paper, this
proposed step-scaling method works well and offers a feasible
approach to heavy quark calculations with accurately controlled
finite lattice spacing errors.  However, unless we can move beyond
the quenched approximation used here, this method will be of only
limited utility.  Using this method for full QCD will, of course,
be more computationally demanding because each of the two sets of
lattice configurations generated for this matching process must be
obtained from a full, dynamical simulation including the quark
determinant.  However, such an approach could become prohibitively
expensive if the value of the light dynamical quark masses,
$m_{\rm light}a$, must to decrease toward zero with decreasing $a$.

Fortunately, such small dynamical quark masses are not required
in this step-scaling approach.  The combination of the usual gauge
and light-quark action together with the effective lattice action of
Eq.~\ref{eq:RHQ} defines a complete physical theory, including the
properties of heavy quarks, that is unambiguously specified at short
distances $\lambda$ with $a \ll \lambda \ll 1/m_{\rm light}$.
Recall that in the continuum such a local field theory is typically
defined in a mass- and volume-independent fashion.  Short-distance
renormalization conditions are imposed to fix the theory in a manner
that is insensitive to quark masses and space-time volumes.
Similarly our fine-lattice theory, viewed as a function of the bare
input lattice parameters also defines such a mass- and
volume-independent theory.

Given sufficient computer power, the implications of this theory could
be worked out on arbitrarily large spatial volumes and for arbitrarily
small masses.  The results would be well-defined functions of the
bare input parameters which would require no adjustment as the quark
mass and spatial volume were varied.  The lattice spacing could be
determined in physical units by comparing to $\Lambda_{\rm QCD}$ as
determined from a vertex function at short distances with the light
quark masses having a negligible effect.

Replacing the standard light quark action appropriate for our finest
lattice with the improved lattice action of Eq.~\ref{eq:RHQ} does not
change this situation.  The parameters in the heavy quark action
could be evaluated or renormalized by examining Green's functions
evaluated at non-exceptional momenta without infra-red or light-quark
mass sensitivity~\cite{Lin:2003bu,Lin:2004ht}.  We would still be
working with a short-distance-defined field theory that will give
meaningful predictions as a function of lattice volume and light
quark mass.  Thus, when comparing two such effective theories defined
at two different lattice spacings we are free to use any lattice size
$L$ and dynamical quark mass $m_{\rm light}$ we find convenient
provided $L \gg a$ and $m_{\rm light} \ll 1/a$.  In fact, if
$m_{\rm light}$ is sufficiently small that it does not effect the
finite-volume heavy-quark spectrum being compared, we need not even
use the same quark mass in the two calculations being compared!  Of
course, this is likely a quark mass that is expensively small and a
better strategy is to work with sufficiently small spatial volume and
sufficiently heavy dynamical quark mass that they do effect the
quantities being matched and must be given equivalent physical
values in each of the calculations being compared.

We conclude that employing the procedure developed here in full
QCD, while difficult, is practical and well within the reach of
present computer resources.  Just as our step-scaled lattice spacing
decreases and we move to increasingly smaller spatial volumes, we
should also move to increasingly heavier quark masses.  In both cases
finite volume and finite dynamical quark mass effects are distorting
the spectrum being compared, but these distortions are entirely
physical and must be accurately described by the effective actions
being compared.

\section{Simulations}
\label{Sec:simulations}

We performed this calculation on a 512-node partition of the
QCDOC machine located at Columbia University.  We used the Wilson
gauge action since for this case the relation between lattice
coupling and lattice spacing has been thoroughly
studied~\cite{Guagnelli:1998ud,Necco:2001xg}.

The gauge configurations were generated using the heatbath method
of Creutz~\cite{Creutz:1980zw}, adapted for $SU(3)$ using the
two-subgroup technique of Cabibbo and Marinari~\cite{Cabibbo:1982zn}.
The first 20,000 sweeps were discarded for thermalization and
configurations thereafter were saved and analyzed every 10,000
sweeps.  We examined the auto-correlation between configurations
for both the standard 4-link plaquette and the hadron propagators
evaluated at a time separation of 12 lattice units.  Here we use the
standard autocorrelation function $\rho(t)$ as
\begin{eqnarray}\label{eq:AutoCorr}
 \rho(t) = \frac{1}{N_{tot} - t}\sum_{j}^{N_{tot} -t} \left(O(j)
  - \overline{O}\right)\left(O(t+j) - \overline{O}\right)
\end{eqnarray}
and identify the autocorrelation time as the size of the region
near $t=0$ in which $\rho(t) \ne 0$.

\begin{figure}[hbt]
\includegraphics[width=\columnwidth]{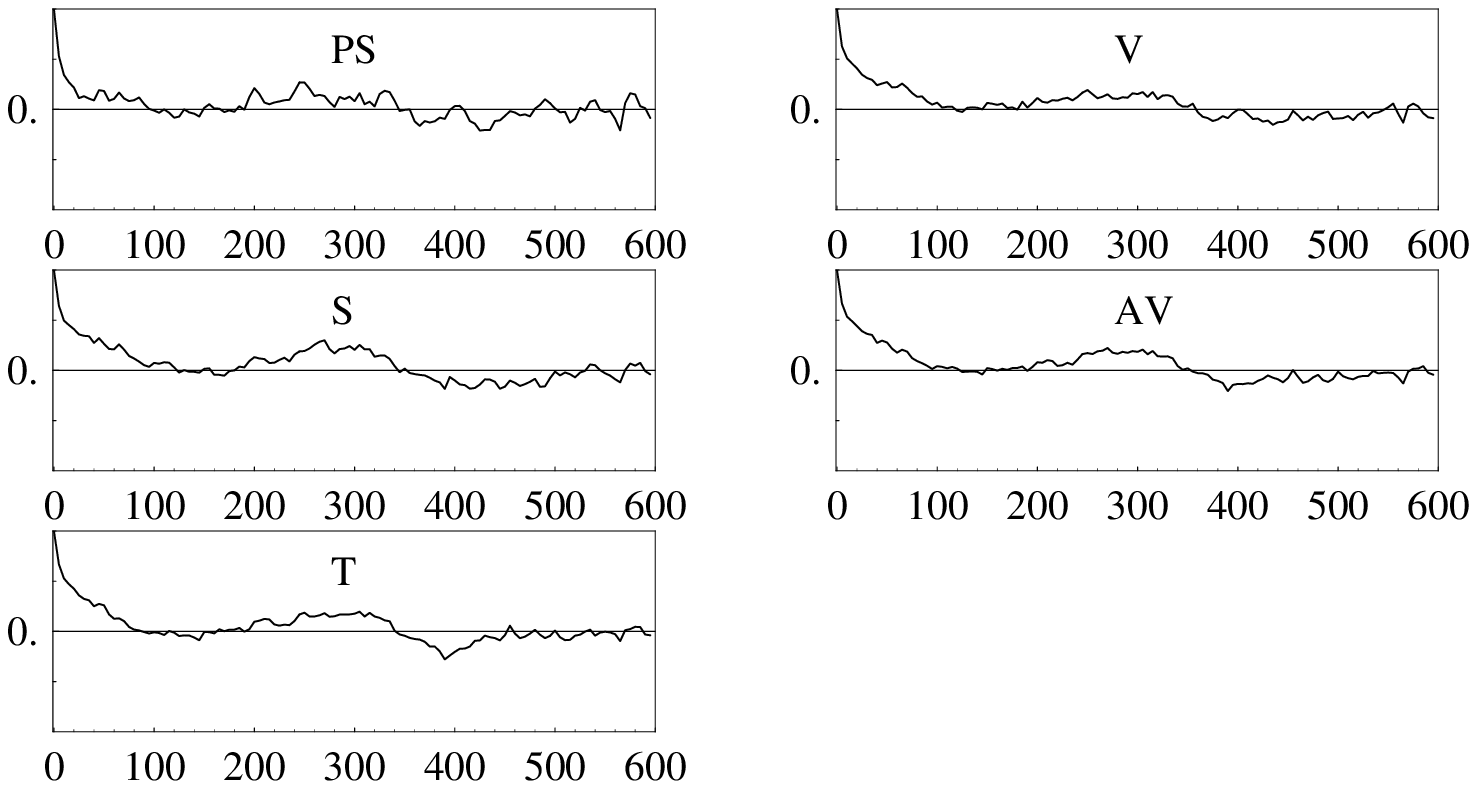}
\includegraphics[width=\columnwidth]{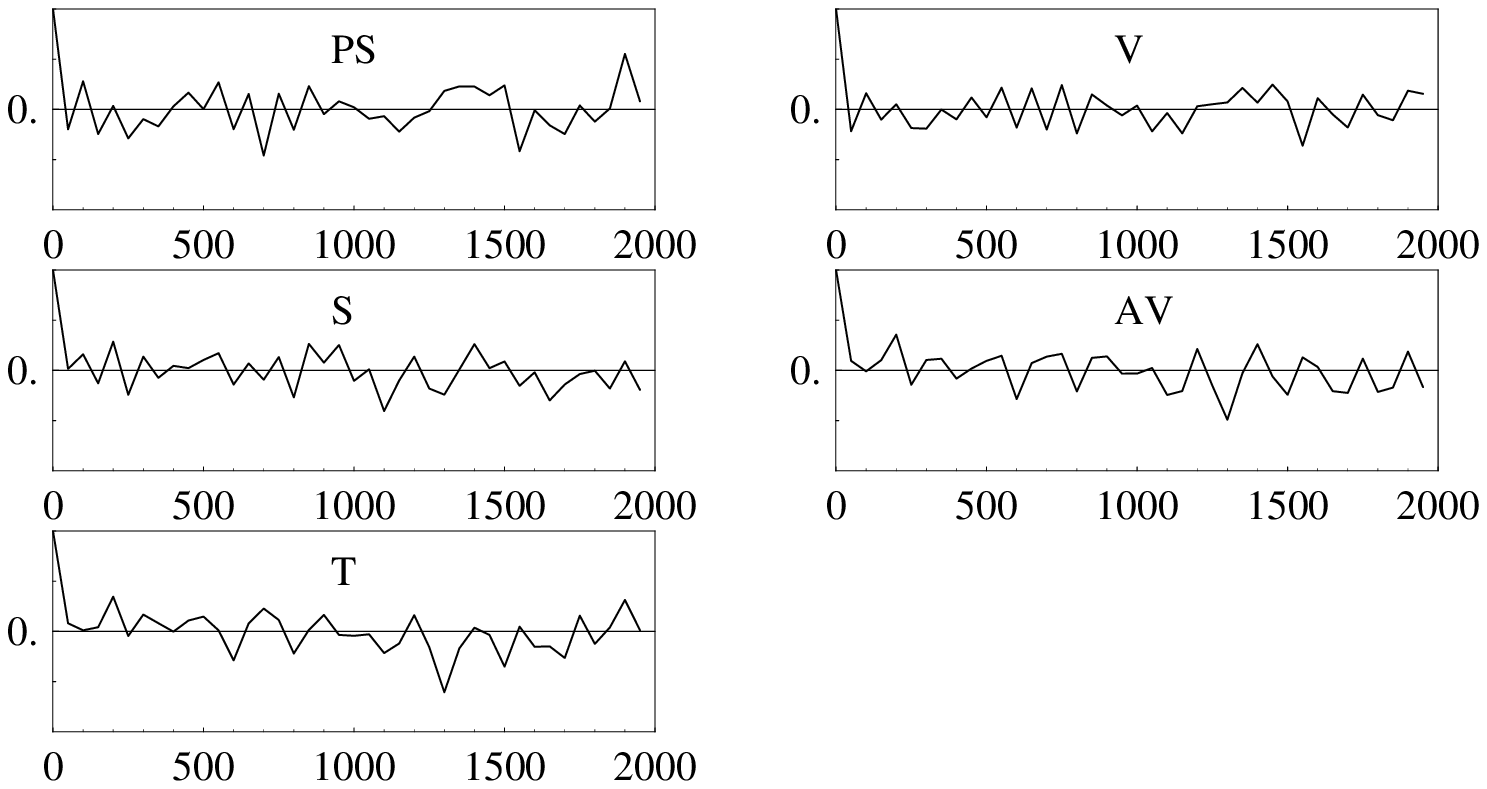}
\caption{\small\small  The auto-correlation function for five
different heavy-heavy meson propagators evaluated at a source-sink separation
of twelve lattice units with $\beta =6.638$ and a $24^3\times48$
space-time volume.   In the top graph the propagators were calculated
on every 5th configuration while in the bottom graph the propagator
measurements were separated by 50 sweeps.  This suggests that our
separation of 10,000 sweeps between measurements ensures that
they will be uncorrelated.}
\label{fig:PropAutoCorr} \vskip -3mm
\end{figure}

For the case of the plaquette (which was calculated every sweep) we
found an auto-correlation time of approximately 3 sweeps.  We studied
the propagator correlations using two $24^3 \times 32$, $\beta=6.638$
test calculations: in the first, the propagators were computed on 120
configurations, separated by 5 sweeps and in the second on 40
configurations separated by 50 sweeps.  The resulting correlations
functions for five different hadron propagators evaluated at a
temporal source-sink separation of twelve lattice units are shown in
Fig.~\ref{fig:PropAutoCorr}.  Auto-correlation on the scale of 100
sweeps can be seen in the data sampled every five sweeps.  Essentially
no autocorrelation can be seen for the propagators sampled every
fifty sweeps.  Since we used a total of 100 such lattice configurations,
separated by 10,000 sweeps, for all of the quantities discussed in this
paper, we will assume that quantities calculated on different
configurations are statistically independent.

\subsection{Lattice scale}
\label{subsec:lattice_spacing}

Four different quantities with a meaningful physical scale
enter each of the two lattice calculations that must be matched
at a given stage in our step-scaling procedure.  Most familiar
is the distance scale determined by the static quark
potential or the chiral limit of the light hadron spectrum.
This is an important physical scale that will influence
even small-volume, heavy quark results.  As is conventional,
we will refer to this quantity as the ``lattice spacing''.
We will find it convenient to determine this from direct
calculation of the static quark potential.  The other three
scales are the lattice volume, and the masses of the light and
heavy quarks.  (In our discussion of heavy-light systems
we will ignore the strange quark and work with degenerate
up and down quarks.)

Of course, the lattice spacing, expressed in physical units,
is also important since it gives us a direct idea of the
size of the discretization errors which we are trying to
control.  For this purpose we don't need great precision
(something that cannot be achieved in a quenched calculation
under any circumstances).  We will determine the lattice
spacing in physical units from the static quark potential
evaluated at an intermediate distance to yield the Sommer
scale~\cite{Guagnelli:1998ud} which is then determined from
a phenomenological, static quark potential model.   While
the later is not precisely defined, this method has the
advantage that it uses only pure gauge theory without any
fermion action being involved.  We could get a physical
value for the lattice scale from the pion decay constant
$f_\pi$ or the rho meson mass but these are more difficult
to calculate and may be more sensitive to finite volume
and other systematic errors.

Our strategy for choosing the parameters to be used on the
fine lattice and then finding physically equivalent parameters
to be used on the coarser lattice proceeds as follows.  We
first decide on a target ratio of lattice scales determined
by the ratio of lattice volumes that we intend to use.  In
the present case a ratio of $3/2$ is implied by our choice of
$24^3$ and $16^3$ lattice volumes.  Second, we choose the bare
lattice coupling on the fine lattice to insure that the fine
lattice spacing is sufficiently small (here chosen to be
$1/a= 5.4$ GeV).   The stronger coupling must then give a lattice
spacing larger by a factor of 3/2 than the fine value or
$1/a=3.6$ GeV.  While for a dynamical QCD calculation, this
would require considerable numerical exploration, for a quenched
calculation with the Wilson gauge action, we can simply refer
to extensive earlier work.

Next we choose the light quark mass to be used on the fine
lattice as sufficiently light that the heavy-light mesons
being studied will be involve a different momentum
scale than do the heavy-heavy mesons but not so light as to
unreasonably increase the computational cost.  For the
calculations reported here we used the domain wall formalism
for the light quarks and chose the mass $m_f a = 0.02$,
one-tenth of the 0.2 heavy quark mass.  The light quark mass
to be used on the coarse lattice is determined by requiring
that the light-light pseudo-scalar meson have a mass 3/2 times
larger than that found on the fine lattice when measured in
lattice units.

Finally the heavy quark mass on the fine lattice is estimated
to correspond to the bare charm quark mass.  While in the present
calculation we have used the single value $m_f a = 0.2$, a
complete calculation will likely require one or two more masses
so that a final interpolation/extrapolation can be done to make
the physical charmed hadron mass agree with experiment.  The
heavy quark mass on the coarse lattice is one of the three
heavy-quark parameters whose determination is discussed in the
next section.

Let us now discuss the choice of lattice scale in more detail.
The static potential is expected to have the following form
\begin{eqnarray}\label{eq:PotentialForm}
V(R) = C - \frac{\alpha}{R} + \sigma R
\label{eq:static_pot_th}
\end{eqnarray}
where R is the separation between the static quarks.  The scale
implied by the heavy quark potential is often specified using
the Sommer parameter $r_0$ which is defined by the condition
\begin{eqnarray}\label{eq:SQPr0}
- R^2 \frac{\partial V(R)}{\partial R} |_{R=r_0} = 1.65.
\end{eqnarray} %
This is appropriate on standard size lattices for bare
couplings in the range $\beta=6/g^2 \le 6.57$.  For weaker
couplings, $\beta > 6.57$ one uses a second, smaller distance
scale, $r_c$ defined by
\begin{eqnarray}\label{eq:SQPr1}
R^2 F(R) |_{R=r_c} = 0.65
\end{eqnarray}
where $\frac{r_c}{r_0} = 0.5133(24)$~\cite{Necco:2001xg}.  While
it may be problematic in a quenched calculation, we can attempt to
determine $r_0$ from a phenomenological potential model, which
gives $r_0^{-1}=0.395$~GeV.

Reference~\cite{Necco:2001xg} gives predictions for the resulting
lattice spacing when the coupling $\beta$ of Wilson gauge action
is in the range $5.7~\le\beta~\le6.92$:
\begin{eqnarray}\label{eq:SommerScale}
\ln(a/r_0) = -1.6804 - 1.7339(\beta-6) + 0.7849(\beta-6)^2 -
0.4428(\beta-6)^3.
\end{eqnarray}
With the help of Eq.~\ref{eq:SommerScale}, we can locate the
$\beta$ values needed to achieve the desired cutoff scales and
fine tune it later as necessary. As our final choices, we have
$\beta = 6.638$ for the $a^{-1}=5.4$~GeV lattice and $\beta =
6.351$ for the $a^{-1} = 3.6$~GeV one.

Since the comparison of lattice scales between our two simulations
is fundamental to this matching program, we have carried
out additional calculations to make sure that the lattice spacing
is correctly selected.  This requires a direct calculation of
the static quark potential on our lattice configurations.

Recall that the static quark potential can be extracted from
the ratio of Wilson loops:
\begin{eqnarray}\label{eq:StaticQuarkPot}
V(\vec r) = \log \left[\frac{\langle W(\vec r,t) \rangle}{\langle
W(\vec r, t+1)\rangle}\right]
\end{eqnarray}
where $\langle\ldots\rangle$ denotes an average over gauge
configurations.  In order to improve the signal and to extract
the potential $V(r)$ from smaller time separations, we smear
the gauge links in the spatial directions according to
Ref.~\cite{Albanese:1987ds}
\begin{eqnarray}\label{eq:APEsmear}
U_k(n) \rightarrow P_{{\rm SU}(3)}\left[U_k(n) + c_{\rm smear}\sum_{\rm l \neq
k}U_{l}(n)U_{k}(n + \hat{\rm{l}}) U_{-l}(n +
\hat{l} + \hat{k} ) \right]
\end{eqnarray}
where $k$ and $l$ each indicate a spatial direction, $P_{{\rm SU}(3)}$
is an operator that projects a link back to an ${\rm SU}(3)$
special unitary matrix, $c_{\rm smear}$ is the smearing
coefficient (set to 0.5 in our case) and the smearing
procedure is performed $n_{\rm smear}$ times. More details
regarding the algorithm can be found in
Ref.~\cite{Hashimoto:2004rs,Aoki:2004ht}. In our calculation we
found good results for $n_{\rm smear}=180$ for the fine lattice
and $n_{\rm smear}=60$ for the coarse one.

\begin{figure}[hbt]
\begin{center}
\vskip 3mm
\includegraphics[width=0.7\columnwidth]{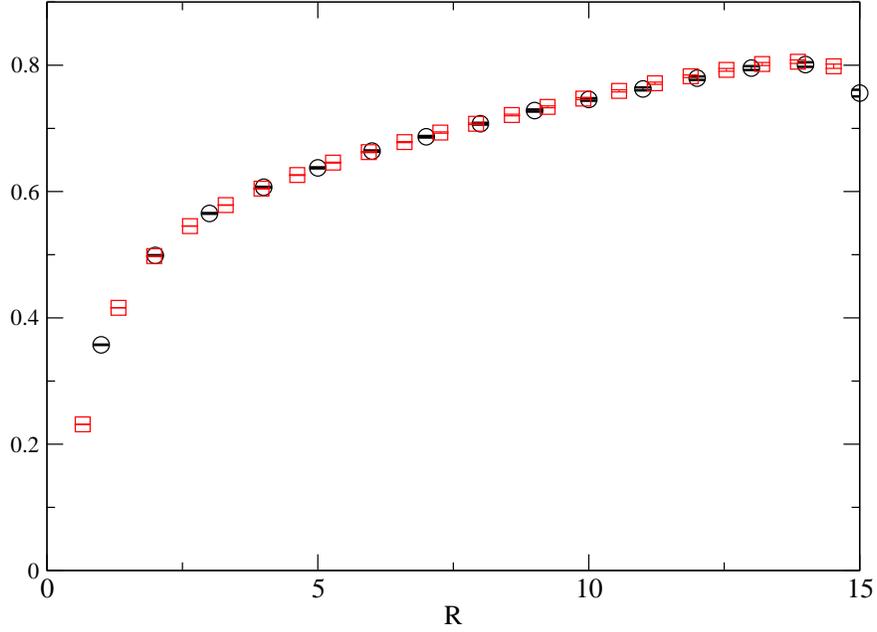}
\end{center}
\caption{The static quark potential calculated on both the coarse
and fine lattices.  The static quark potential values computed on
the $\beta = 6.351$ lattice are shown as circles.  The squares mark
the static quark potential from the $\beta= 6.638$ lattice scaled
by the fitted ratio of lattice spacings $\lambda=1.51$ and shifted
by a constant.   The agreement between these two different sets of
points gives good evidence that the ratio of lattice spacings
between these two $\beta$ values is the desired 3/2.}
\label{fig:static_ratio} \vskip-3mm
\end{figure}

While we did determine the two scale standards $r_0$ and $r_c$
individually for both of our lattice spacings, our lattice volumes
are somewhat small to permit a comparison with infinite volume
results. We therefore also determined the ratio of lattice spacings
without using the Sommer scale by directly comparing the potentials
computed on our two sets of lattice configurations using the
relation
\begin{eqnarray}\label{eq:PotentialRatio}
V_1(n)= V_2(n/\lambda)/\lambda + C' \nonumber
\label{eq:static_ratio}
\end{eqnarray}
where $\lambda={a_2}/{a_1}$ is the ratio of the two lattice spacings.
We first fit the static potential on the fine lattice to the form
given in Eq.~\ref{eq:static_pot_th}, determining the parameters $C$,
$\alpha$ and $\sigma$.  Next we scaled the resulting fitted function
according to Eq.~\ref{eq:static_ratio} and adjusted the parameters
$\lambda$ and $C'$ in that equation to obtain the best fit to the static
potential measured on the coarse lattice.  Figure~\ref{fig:static_ratio}
shows a comparison of the potential determined from $\beta=6.351$
configurations and a scaled and shifted version of the
$\beta=6.638$ potential.  The agreement is excellent and this
procedure gives an independent value for the lattice spacing ratio
of $1.51(2)$, which agrees with what we wanted.

\subsection{Domain-Wall Fermions}
\label{subsec:DWF}

We will now briefly describe the domain wall fermion calculations
that were used for the heavy quark on the finest lattice
and the light quarks on both lattices.  The domain wall Dirac operator
can be written as
\begin{equation}\label{eq:Ddwf}
  D_{x,s; x^\prime, s^\prime} = \delta_{s,s^\prime}
    D^\parallel_{x,x^\prime} + \delta_{x,x^\prime} D^\bot_{s,s^\prime}\nonumber
\end{equation}
\begin{eqnarray}\label{eq:DdwfParallel}
D^\parallel_{x,x^\prime} & =&
  {1\over 2} \sum_{\mu=1}^4 \left[ (1-\gamma_\mu)
  U_{x,\mu} \delta_{x+\hat\mu,x^\prime} + (1+\gamma_\mu)
  U^\dagger_{x^\prime, \mu} \delta_{x-\hat\mu,x^\prime} \right]
+ (M_5 - 4)\delta_{x,x^\prime} 
\end{eqnarray}
\begin{eqnarray}\label{eq:DdwfPerp}
D^\bot_{s,s^\prime}
     &=& {1\over 2}\Big[(1-\gamma_5)\delta_{s+1,s^\prime}
                 + (1+\gamma_5)\delta_{s-1,s^\prime}
                 - 2\delta_{s,s^\prime}\Big] \nonumber\\
     &-& {m_f\over 2}\Big[(1-\gamma_5) \delta_{s, L_s-1}
       \delta_{0, s^\prime}
      +
      (1+\gamma_5)\delta_{s,0}\delta_{L_s-1,s^\prime}\Big].
\end{eqnarray}
where the fifth-dimension indices $s$ and $s^\prime$ lie in the
range $0 \le s,s^\prime \le L_s-1$, $M_5$ is the five-dimensional
mass and $m_f$ directly couples the two walls at $s=0$ and $s=L_s-1$.
It is related to the physical mass of the four-dimensional fermions.

The $M_5$ parameter is optimized by the choice of
$M_5 \approx 1 - m_{\rm crit}$, where $m_{\rm crit}$ is the
critical value of the mass for the 4-dimensional Wilson fermion
action.  This quantity has been calculated perturbatively up to
one-loop level for the Wilson gauge action and either the
Wilson\cite{Follana:2000mn} or SW\cite{Panagopoulos:2001fn}
fermion actions. In the quenched approximation, for Wilson fermions
and with our choices of gauge coupling, we find
$m_{\rm crit} = -0.495$ at $\beta = 6.638$, and
$m_{\rm crit} = -0.522$ at $\beta = 6.351$.  Therefore, we use
$M_5 = 1.5$ in the DWF action for both our $\beta$ values.

The DWF action is $O(a)$ off-shell improved due to the
preservation of chiral symmetry, and no further improvement in the
action or quark fields is performed. The chiral symmetry breaking
can be measured from the residual mass, which can be computed
from the ratio
\begin{eqnarray}\label{eq:Mres}
am_{\rm res} =\frac{\sum_x \langle
J_{5q}^a(\vec{x},t)\pi(0)\rangle}{\sum_x \langle
J_{5}^a(\vec{x},t)\pi(0)\rangle},
\end{eqnarray}
provided $t \gg a$.  Here $J_{5q}$ is a pseudoscalar density
located at the midpoint of the fifth dimension.
The residual mass has been thoroughly studied, for example, in
Ref.\cite{Aoki:2002vt} for various values of $\beta$, $L_s$ and
$M_5$.  Those results suggest that the $m_{\rm res}$ values for
each of our lattice configurations are much smaller than the
0.00124 value determined at $\beta=6.0$ with lattice volume
$16^3\times32\times16$ and $M_5=1.8$.  This indicates that chiral
symmetry breaking is small and ignoring the contribution of
$m_{\rm res}$ in the heavy quark sector will have an effect
smaller than 0.5\%.

\begin{figure}[hbt]
\includegraphics[width=\columnwidth]{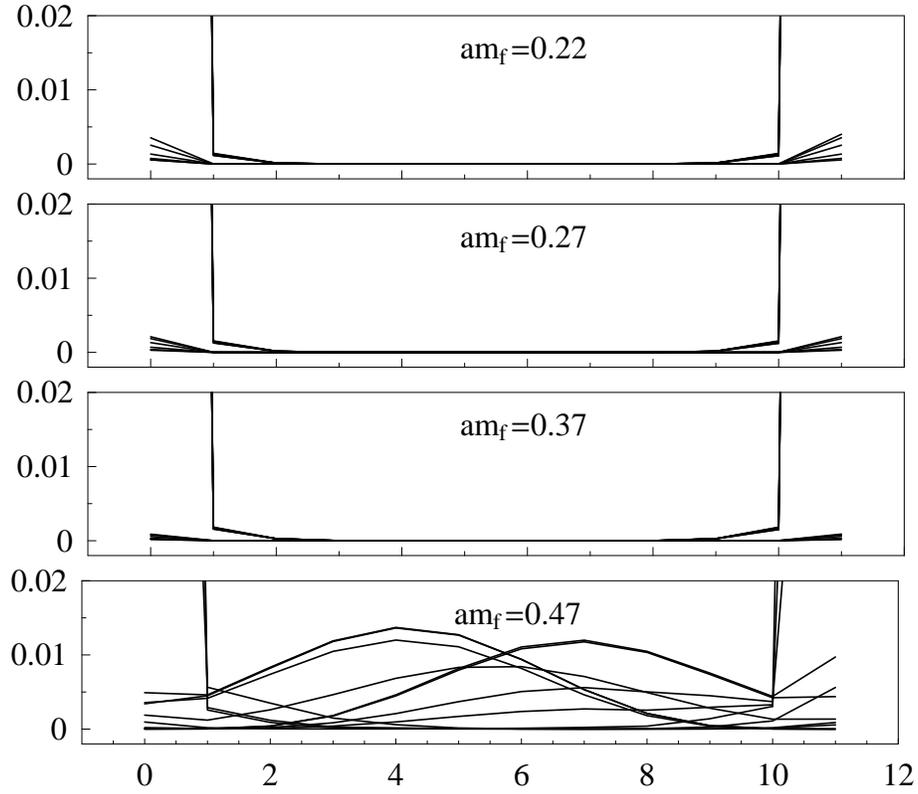}
 \caption{The dependence on the fifth dimension, $s$, of the space-time
sum of the modulus squared of the nineteen lowest-lying eigenvectors
of the hermitian domain-wall Dirac operator: $\sum_x|\Psi_{x,s}|^2$.
This is shown for a Wilson quenched gauge configuration with
$\beta = 6.638$, $V = 16^4$, $L_s = 12$, $M_5 = 1.8$, and
$m_f \in \{0.22, 0.27, 0.37, 0.47\}$. The unphysical, propagating
states are seen only for $m_f > 0.37$.}
\label{fig:dwfEigenLog}\vskip -3mm
\end{figure}

However, there is a limitation to using large values of $m_f$
with DWF.  Recall that there are two types of eigenvectors of
the hermitian DWF Dirac operator: propagating and
decaying states\cite{Christ:2004gc,Liu:2003kp}. The former,
unphysical states have non-zero $5^{th}$-dimension momenta and
large Dirac eigenvalues around $1/a$.  The ``decaying'' states
are bound to the walls of the $5^{\rm th}$ dimension and are the
physical states corresponding to the four-dimensional Dirac
eigenstates in the continuum limit.  The gap between these two types
of states is controlled by the domain-wall height $M_5$. However,
as $m_f$ increases, the eigenvalues of the physical states
increase while those of the propagating states do not.  Thus, we
must be careful to avoid the situation in which the states with
the smallest eigenvalues are dominated by these unphysical states.
Therefore, a careful check on the lowest eigenvalues for the
target $m_f$ being used to simulate the heavy quarks on the
fine lattice is needed.   Figure~\ref{fig:dwfEigenLog} shows the
5-dimensional eigenfunction, averaged over 4-dimensional
space, $\sum_x|\Psi_{x,s}|^2$ as a function of the
$5^{\rm th}$-dimensional coordinate $s$ for the lowest nineteen
eigenvalues with various $m_f$: 0.22, 0.27, 0.37, 0.47.  As
can be seen in the figure, these first nineteen eigenfunctions
appear to be physical states bound to the 4-dimensional wall
for the first three mass values.  However, $m_f=0.047$ is
sufficiently large that propagation into the 5-dimension can
be clearly seen.  We conclude that our $m_f=0.2$ for the heavy
quark is safe, well below the region where such unphysical
states arise.

\subsection{Spectrum Measurements}
\label{subsec:SpectrumMeasurement}

In order to get good signals for the heavy quark states of interest
for relatively small time separations, a smeared wavefunction source
is used for the heavy quark (but not the light quark).  Here, we
adopt the Coulomb gauge-fixed hydrogen ground-state wavefunction:
\begin{eqnarray}
\Psi_{gnd}(r) &=& e^{-r/r_0}
\label{eq:SmearFunc}
\end{eqnarray}
as the source of the heavy fermion(s) and use a point source for the
light fermion (if any).  At the sink the two propagators are evaluated
at the same point and the resulting gauge invariant combination
summed over a 3-dimensional plane at fixed time, with a possible momentum
projection factor.  An optimized radius, $r_0$, was chosen in the
fashion suggested in Ref.~\cite{Boyle:1999gx}. Table~\ref{tab:meson_op}
lists all the local meson operators used in our calculation.

\begin{figure}[hbt]
\includegraphics[width=\columnwidth]{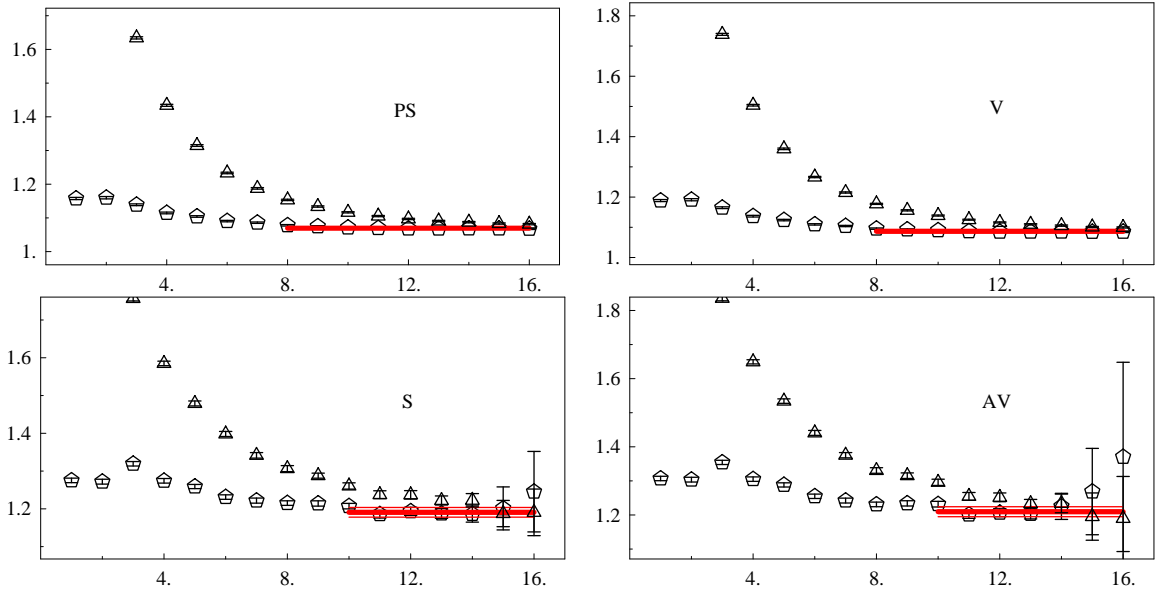}
\caption{A comparison of effective mass plateaus between point
and smeared sources in the heavy-heavy sector.  Triangles
denote the point-point source, and pentagons denote the
point-smeared heavy meson correlators.  Clearly, the plateaus have
been improved by the smearing.  However, a double-cosh fit to the two
distinct wavefunction sources might help us determine the ground state
energy even more accurately.}
\label{fig:effMass}\vskip -3mm
\end{figure}

Figure~\ref{fig:effMass} shows how the plateau in the effective
mass plot improves between a point and smeared source.  The smeared-source
meson plateaus are much better those of the local source, even for the
scalar and axial-vector mesons.

To constrain the space-time asymmetry parameter $\zeta$, we also
computed the pseudo-scalar meson energy in the heavy-heavy sector
for the three lowest on-axis momenta:
$\frac{2\pi}{L} (0, 0, 0)$, $\frac{2\pi}{L} (0, 0, 1)$, and
$\frac{2\pi}{L} (0, 1, 1)$, where $L$ is the spatial lattice size.
The dispersion relation may be expanded in momentum as
\begin{eqnarray}\label{eq:DisFunc}
E(p) & = & m_1 + \frac{p^2}{2 m_2} + O(p^4).
\end{eqnarray}
As we will see, requiring the ratio of static to kinetic mass,
$m_2/m_1=1$ is useful for determining the coefficient $\zeta$.

\subsection{Parameters}
\label{subsec:parameters}

Table~\ref{tab:fixedPar} lists the fixed parameters used
throughout this matching stage for the fine and coarse lattices.
The heavy quark mass was set to approximate that of the charm
mass and the light quark mass was chosen ten times smaller.  The
lattice spacing ratio between these two lattices is 1.5.  The
domain wall fermion parameters used on the fine lattice have been
carefully studied and we find no unphysical states in the chosen
mass range as discussed in previous section.  Table~\ref{tab:FineData}
shows the hadron mass spectrum computed on the fine lattice.  As
can be seen, $m_1/m_2=1.02(2)$ is consistent with 1, indicating that
heavy quark discretization effects using domain wall fermions are
small.  One expects that the light quark mass on the coarse lattice
should be 0.03 and the data for the light-light spectrum with this choice of
light quark mass is listed in Table~\ref{tab:CoarseDataLight}.
As one can see from Table~\ref{tab:CoarseDataLight}, the light-light
meson spectra on the coarse and fine lattices agree when compared
in the same units, indicating that the light quark mass is well tuned
on the coarse lattice.

\begin{figure}[hbt]
\includegraphics[width=\columnwidth]{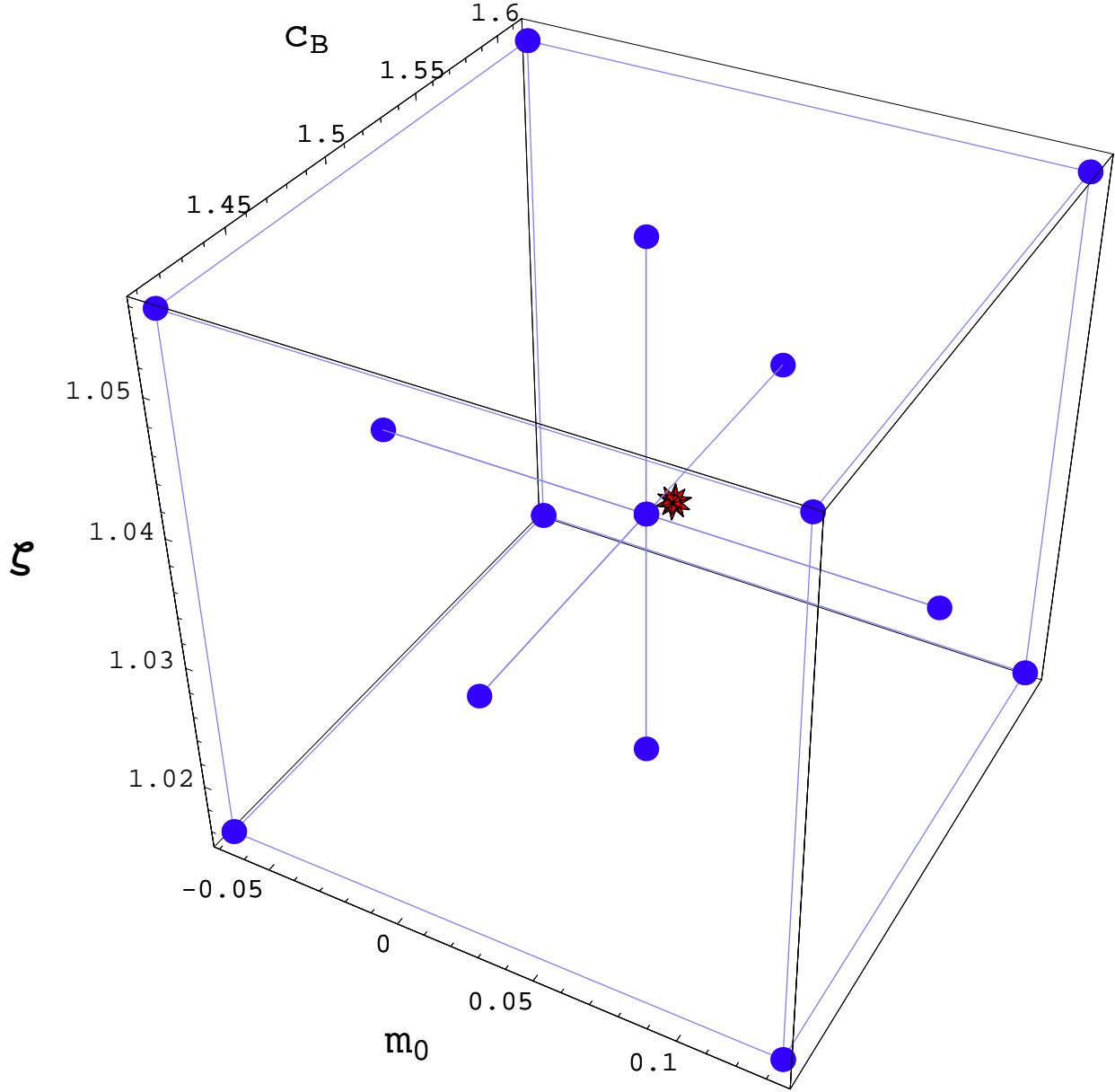}
\caption{The distribution in the 3-parameter space
($m_0$, $c_B$, $\zeta$) of the 24-set ``Cartesian'' data. The
center circular point is set \#14, and the points around it are
sets\#43-66. The starred point represents the final matching
coefficients determined in Sec.~\ref{subsec:ThreeParAction}.}
\label{fig:24CartesianSet}\vskip -3mm
\end{figure}

A complete list of the parameter sets used for the RHQ action on the
coarse lattice is given in Table~\ref{tab:AllCoarsePara}.  The first
42 sets of data were initial trials chosen to give good coverage in
parameter space.  In order to perform a more systematic analysis, described
in Sec.~\ref{Sec:Analysis}, we also collected a ``Cartesian'' set
(sets~\#43-\#66) chosen close to the desired fine lattice measurements.
These 24 data sets are centered around set~\#14.  The range of each
parameter in this Cartesian data set was selected so that within that
range the estimated difference between a linear and quadratic fit
would be less than 5\% as expected from an examination of the first
42 parameter sets.  This yields a region that is close to reproducing
the target fine data and in which a linear approximation should be
good: $m_0 = 0.0328 \pm 0.1$, $c_B = 1.511 \pm 0.1$, $c_E = 1.538 \pm 0.3$
and $\zeta = 1.036 \pm 0.02$, which is shown in Figure~\ref{fig:24CartesianSet}.

The 24-set ``Cartesian'' data will allow us to calculate the first and
second derivatives directly from the measurements.  Note that we have
more measurements than we actually need.  This provides additional
checks on our method and the validity of the scaling of physical
quantities between the coarse and fine lattices.  We expect that the
total number of data sets that we will use for next step of matching
will be dramatically reduced.

Using the methods described in Sec.~\ref{subsec:SpectrumMeasurement},
we have measured the pseudoscalar~(PS), vector~(V), scalar~(S) and
axial-vector~(AV) mesons in the heavy-heavy system, and PS and V
in the heavy-light system. We use combinations of the masses to
try to simplify their dependence on the coefficients of the RHQ
action. For the heavy-light system, we use the spin-average and
hyperfine splitting; for the heavy-heavy system, we use these and
also include the spin-orbit average, spin-orbit splitting and
the ratio of $m_1/m_2$.  The resulting values for these quantities
for each of the 66 data sets are given in Table~\ref{tab:AllCoarseData}.

\section{Analysis and Results}
\label{Sec:Analysis}

The final step in this matching procedure is to determine the
parameters in the heavy quark action of Eq.~\ref{eq:RHQ},
$\{m_0,c_B, c_E, \zeta\}$, that will yield the seven quantities
measured on the coarse lattice which agree with those determined on
the fine lattice.  Of course, this might be done by ``trial and
error'' and, as can be seen by scaling the numbers in
Table~\ref{tab:AllCoarseData}, data set \#14 comes very close to such
a result.  However, to fully understand this step scaling method
(for example to properly propagate errors) it is important to learn
in detail how the measured spectra depend on these input parameters.

As a starting point, we will attempt to use a subset of our parameter
space chosen so that the resulting coarse lattice hadron masses are
well fit by a simple linear dependence on the heavy quark parameters:
\begin{eqnarray}\label{eq:FitCoarse}
Y^n = A + J \vdot X^n,
\end{eqnarray}
where $n$ labels the parameter set while $X$ and $Y$
are 4-dimensional and 7-dimensional column vectors made up of the four
input heavy-action parameters and the seven computed masses or mass
ratios, respectively:
\begin{equation}
X = \left(\begin{array}{c} m_0 \\ c_B \\ c_E \\ \zeta \end{array}\right) \quad
Y = \left(\begin{array}{c} m^{hh}_{sa} \\ m^{hh}_{hs} \\ m^{hl}_{sa} \\
                     m^{hl}_{hs} \\ m^{hh}_{soa} \\ m^{hh}_{sos} \\ m_1/m_2 \end{array}\right).
\label{eq:coarse_def}
\end{equation}
The quantities $A$ and $J$ are a 7-dimensional column vector and a
$7 \times 4$ matrix which represent the constant and linear terms in
our linear approximation.  (In most of the discussion to
follow we will work with all seven measured quantities.  However, if
this number is decreased, the vectors $Y$, $A$ and the matrix $J$
will shrink appropriately.)

Given a specific group of $N$ of our data sets, $X^{n_i}|_{1 \le i \le N}$,
we can determine the quantities $A$ and $J$ by minimizing an appropriate
$\chi^2$ for such a linear fit:
\begin{eqnarray}\label{eq:ChiCoarse}
\chi_{\cal C}^2 = \sum_{i=1}^N
               (A + J \vdot X^{n_i} - Y^{n_i})^T
 \vdot W_{\cal C}^{-1} \vdot (A + J \vdot X^{n_i} - Y^{n_i}).
\end{eqnarray}
Here $W$ is a $7 \times 7$ matrix representing a choice of correlation
matrix.  In the results that follow we will use
\begin{eqnarray}
(W_{\cal C})_{d,d'} = \sum_{i=1}^N \langle
                   \Bigl(Y^{n_i}_d    -  \overline{Y}^{n_i}_d\Bigr)
                   \Bigl(Y^{n_i}_{d'} - \overline{Y}^{n_i}_{d'}\Bigr) \rangle
\label{eq:Wc}
\end{eqnarray}
where $\langle ... \rangle$ represents an average over the 100 jackknife blocks
obtained by omitting one of the 100 measurements with $Y^{n_i}_d$ the
result for that jackknife block and $\overline{Y}^{n_i}_d$
the corresponding average.  Replacing $W_{\cal C}$ by the simpler, uncorrelated
error matrix $(W_{\cal C}^\prime)_{d,d'} = \sum_{i=1}^N\delta_{d,d'} \sigma^{n_i}_d$ had
little effect on the final results where $\sigma^i_d$ is the usual squared error
on the measured quantity $Y^i_d$.  Determining the $A$ and $J$
which minimize $\chi_{\cal C}^2$ is straight-forward because this is a
quadratic function of these 35 numbers and the minimum can be obtained
by solving 35 linear equations.  Typically these 35 equations are quite
regular, with a stable solution even if only a relative few of our data
sets are used.

The use of linearity to determine the desired matching heavy quark
parameters is reasonable if we are working in a region that is close
to the right choice for those parameters.  Once we have determined the
matrix $J$ and vector $A$, we can solve for the coefficients $X_{\cal C}$
that will yield meson masses equal to those found on the fine lattice,
$Y_{\cal F}$.  Here we add the subscripts $\cal C$ and $\cal F$ to
indicate our estimates for the physical coarse-lattice parameters
($X_{\cal C}$) and the coarse-lattice masses ($Y_{\cal F}$), scaled from
those calculated using the fine lattice.

Again we minimize a quantity $\chi_{\cal F}^2$, similar to that given in
Eq.~\ref{eq:ChiCoarse}.  However, the fine-lattice correlation matrix,
$W_{\cal F}$, which appears in the equivalent version of Eq.~\ref{eq:ChiCoarse}
is defined through a modified version of Eq.~\ref{eq:Wc}.  Specifically,
the fine-data analogue of Eq.~\ref{eq:Wc} is used for all but the
seventh row and seventh column, which correspond to the quantity $m_1/m_2$.
Since this must be unity in a relativistic calculation (and is one within
errors for our DWF results), we set $(Y_{\cal F})_6 = 1$ and the
corresponding elements of the correlation $(W_{\cal C})_{d\ne 6,6} =
(W_{\cal C})_{6,d\ne 6} =0$ for $0 \le d \le 6$.  In order that the
resulting correlation matrix be invertible, we arbitrarily set
$(W_{\cal C})_{6,6}=10^{-8}$.  This has the effect of constraining the
coarse-lattice value of $m_1/m_2=1$.  The resulting minimum is again
determine by solving a set of linear equations.  That solution can be
written explicitly as:
\begin{eqnarray}\label{eq:FitFine}
X_{\cal C} & =& \left( J^T\vdot W^{-1}_{\cal F} \vdot J \right)^{-1}
\vdot J^T \vdot W_{\cal F} \vdot \left(Y_{\cal F} - A \right).
\label{eq:X_extract}
\end{eqnarray}

Finally, to determine the error on the resulting heavy quark parameters
$X_{\cal C}$ we add in quadrature two different sources of error.  To
compute the first, we use the average value of the masses $Y_{\cal F}$,
deduced from the fine-lattice calculation, which together with
jackknifed results for $J$ and $A$ gives us the error on $X_{\cal C}$
coming from the statistical fluctuations in the coarse lattice data.
We then estimate the statistical error coming from the fine lattice
calculation by using the average values for $J$ and $A$ in
Eq.~\ref{eq:FitFine} and the jackknifed values of $Y_{\cal F}$ to
determine the resulting fluctuations in the resulting heavy quark
parameters, $X_{\cal C}$ caused by the statistical errors in the
determination of $Y_{\cal F}$.

\subsection{Four-Parameter Action}
\label{subsec:CoarseFourParAnly}

As is suggested by the large number of data sets listed in
Table~\ref{tab:AllCoarsePara}, we had greater difficulty than
expected in determining the four parameters $m_0$, $c_B$, $c_E$
and $\zeta$.   Typically, reasonable choices of a subset of
the parameter sets from the initial group of 42 parameter sets
listed in Table~\ref{tab:AllCoarsePara} gave similar values for
the final heavy quark parameters.  However, the derivative matrix
was typically quite singular and the resulting parameters,
especially $c_E$, not well determined.  In an attempt to make this
process more deterministic, we collected the 24 Cartesian data
sets from which we could determine the matrix of derivatives $J$
from simple differences.  The result for $J$ agreed very well
with that typically determined from the fitting procedure
describe earlier to the less regular parameter choices in our
first 42 data sets.  We conclude that this linear description of
our coarse lattice data is a good approximation.  For simplicity,
we present only the results from this final determination of
$J$ and $A$ from the Cartesian data.

Specifically, the twenty four parameter choices within our
Cartesian data set (\#43-\#66) use parameters of the form
\begin{equation}
X^n_i = \overline{X}_i
                        +\sigma(n)_i \Delta_i.
\end{equation}
Here the quantity $\{\sigma(n)_i\}_{0 \le i \le 3}$ determines the
first sixteen parameter choices, where $\sigma(n)_i
=(-1)^{{\rm int}(n/2^i)}$, the expression $\mbox{int}(x)$ represents
the integer part of the number $x$ and the index $n$ varies between
0 and 15.  The four parameter increments $\Delta_0 = 0.1$,
$\Delta_1 = 0.1$, $\Delta_2 = 0.3$ and $\Delta_3 = 0.02$ were
listed earlier and are displayed in Figure~\ref{fig:24CartesianSet}.
The remaining eight data sets use the values $\sigma(16+n)_i=
(-1)^n \delta_{{\rm int}(n/2),i}$ for $n=0,1\ldots 7$.  The quantities
$A$ and $J$ can be directly determined using the following expressions:
\begin{eqnarray}
A_d     &=& \frac{1}{24}\sum_{n=0}^{23}Y^n_d \\
J_{d,i} &=& \frac{1}{9}\sum_{n=0}^{23} \sigma(n)_i \frac{Y^n_d}
           {2\Delta_i}.
\end{eqnarray}

We can then substitute the resulting values of $A$ and $J$ into the
linear relation of Eq.~\ref{eq:FitCoarse} and test this linear
description of our coarse-lattice results for the 24 Cartesian
data sets.  The simplest test of linearity should be $\chi_{\cal C}^2$
of Eq.~\ref{eq:ChiCoarse}.  However, for our 24 sets of seven quantities,
the resulting $\chi_{\cal C}^2/(7\cdot24)$ is $\approx 15$ suggesting this
linear description is poor.  This large value of $\chi_{\cal C}^2$ comes
from the linear prediction of the heavy-heavy spin average masses.  If
these are dropped from the calculation of $\chi_{\cal C}^2$, we obtain
$\chi_{\cal C}^2/(6\cdot24)=1.7$, a much more acceptable value.  Looking
more closely, we find the linear prediction for the heavy-heavy spin
average masses agrees with the calculated value with a fractional
discrepancy of 1-2\% for the 24 data sets.  This is certainly a reasonable
accuracy given the systematic errors in determining these masses from our
lattice calculation.  However, since the statistical error on these
quantities, which is used in our definition of $\chi_{\cal C}^2$,
is of the order of 0.1-0.2\% we should expect these large $\chi_{\cal C}^2$
values.  Thus, we conclude that the linear description of the coarse
lattice results is satisfactory.

Using these results for $J$ and $A$, Eq.~\ref{eq:X_extract} and
the procedure outlined in the previous section to determine the
error we can go on to find the coarse lattice parameters
which describe the fine lattice results:
\begin{eqnarray}
X_{\cal C}^T = \{m_0, c_B, c_E, \zeta\}=
              \{-0.018(100), 1.648(227), 0.957(904), 1.038(23)\},
\end{eqnarray}
where the superscript $T$ indicates the transpose of the column
vector $X_{\cal C}$.  The results for $m_0$ and $\zeta$ are reasonably
accurate.  Note the relative error in $m_0$ should not be determined by
comparing to the central value for $m_0$ which is shifted by the additive
renormalization implied by $m_{\rm crit}$ to be close to zero.  Rather,
one should recognize that this error in $m_0$ corresponds to a 4\%
relative error in $m_{\rm sa}^{hh}$.  However, the errors on $c_B$
and especially $c_E$ are unacceptably large.

In order to better understand these large errors, we now examine
the matrix $J^T \vdot J$.  This matrix is closely related to the
matrix $J^T W_{\cal F}^{-1} J$ which is inverted in Eq.~\ref{eq:X_extract}
to obtain the coarse-lattice parameters.  While the characteristics
of the matrix $J^T W_{\cal F}^{-1} J$ are entirely similar to those
of $J^T \vdot J$ we found it more natural to focus on the simpler
matrix $J^T \vdot J$ whose definition does not depend on a somewhat
{\it ad hoc} choice for the correlation matrix $W_{\cal F}$.

The eigenvalues of the matrix $J^T \vdot J$ are
\begin{eqnarray}
\{ 9.55(15),\;  1.39(10),\;  0.000138(21),\;  0.000037(12)\}
\label{eq:eigenvalues_4}
\end{eqnarray}
with corresponding eigenvectors
\begin{equation}
\begin{array}{rr@{.}lr@{.}lr@{.}lr@{.}ll}
\{& 0&832(4), & -0&1099(6),& -0&1079(7),&  0&532(6)  &\},\nonumber\\
\{&-0&522(7), &  0&062(6), &  0&085(3), &  0&846(4)  &\},\nonumber\\
\{& 0&181(7), &  0&81(7),  &  0&56(10), & -0&003(6)  &\},\nonumber\\
\{& 0&041(23),& -0&57(10), &  0&82(7),  & -0&0156(29)&\}
\end{array}
\label{eq:eigenvectors_4}
\end{equation}
Here the eigenvectors reading top to bottom correspond the
eigenvalues in Eq.~\ref{eq:eigenvalues_4} reading left to right.
The eigenvalues span a range of more than five orders of magnitude
and dramatically decrease between the eigenvectors dominated by the
$m_0$ or $\zeta$ directions and those aligned with $c_B$ or $c_E$.
The smallest eigenvalue corresponds to an eigenvector that has a
large component in the $c_E$ direction which leads to large error
in the $c_E$ coefficient.  (Recall that the components of the
eigenvectors displayed in Eq.~\ref{eq:eigenvectors_4} are arranged
in the order $\{m_0, c_B, c_e, \zeta\}$.

Given the range of quantities measured and the precision of the
results, we were surprised that $c_E$ remains to a large degree
undetermined.  Of course, this is precisely the result that would
be obtained if we were working with a redundant set of parameters.
Thus, we went back and looked carefully at the arguments which
determined this set of ``independent'' parameters and discovered
an additional field transformation that permits $c_E$ to be
transformed into $c_B$.  This result is valid to all orders in
$ma$ and up to errors of order $(a \vec p)^2$.  This theoretical
analysis is presented in the companion paper~\cite{3parAction}.

Here, we will exploit this substantial simplification and use only
the three parameters $m_0$, $\zeta$ and $c_P \equiv c_B=c_E$.  As is
shown in the next section, within this restricted parameter space, the
problem of determining $m_0$, $\zeta$ and $c_P$ from given values
for our seven measured quantities is well-posed and accurate
results for these three parameters can be easily obtained making
our proposed step-scaling, matching procedure quite practical.


\subsection{Three-Parameter Action}
\label{subsec:ThreeParAction}

We will now exploit this simplification from four action parameters
to three and determine those three parameters which give coarse lattice
results agreeing with those found on the fine lattice.  Specifically,
we will use the action in Eq.~\ref{eq:RHQ} but fix $r_s=\zeta$, $r_t=1$
and $c_E=c_B=c_P$ and study the dependence of the seven spectral
quantities making up the vector $Y$ in Eq.~\ref{eq:coarse_def}
on the three parameters $m_0$, $c_P$ and $\zeta$ making up the vector:
\begin{equation}
X^{(3)} = \left(\begin{array}{c} m_0 \\ c_P \\ \zeta \end{array}\right).
\end{equation}
As is shown in Ref.~\cite{3parAction}, a proper, mass-dependent choice
for three parameters will yield on-shell quantities which are accurate
to arbitrary order in $(ma)^n$ with errors no larger than $(a\vec p)^2$.

How does this affect our analysis?  We could, of course, disregard
all of our four-parameter runs and collect an entirely new set of data
with the restriction $c_B=c_E$.  Instead we will exploit the approximate
linearity of much of our four parameter data and interpolate to obtain
what we expect to be a good approximation to the results we would obtain
had we chosen $c_B=c_E$.

Thus, we set $c_{\rm P}=c_B$ and explicitly subtract the deviation
that results from $c_E \ne c_B$ using the matrix of derivatives
$J$ determined in the four-parameter analysis above.  Such an
expansion in $c_B-c_E$ should be especially safe given the very weak
dependence on this difference that we have seen.  Hence the coarse
lattice masses to be used in this three-parameter analysis are obtained
from:
\begin{equation}\label{eq:DataShift}
Y^{(3),n}_d = Y^n_d + J_{d,2}(c_B^n-c_E^n).
\end{equation}
The action parameters corresponding to each of these data sets are
$X^{(3),n}_0 = X^n_0$, $X^{(3),n}_1 = X^n_1$ and $X^{(3),n}_2 = X^n_3$.
The resulting ``three-parameter'' data sets with $1 \le n \le 66$ can
then be analyzed in precisely the same fashion as was done for the
case of four parameters, following the steps taken in
Eqs.~\ref{eq:FitCoarse} through \ref{eq:FitFine}.

Again we use as the center point that data set giving results
closest to the results from the fine lattice, which is
$(X^{(3),14})^T = \{0.0328, 1.511, 1.036\}$ from set~\#14, the ``Cartesian''
data sets~43-66, and obtain
\begin{equation}\label{eq:Xout_3}
(X^{(3)}_{\cal C})^T
    = \{m_0, c_{\rm P}, \zeta\}= \{{ 0.037(26),  1.50(9),  1.029(14)}\}.
\end{equation}
The errors quoted here are statistical and obtained as described in
the beginning of this section by combining in quadrature the errors
coming from the determination of the fine lattice masses and the
statistical uncertainties in determining the coarse lattice
parameters which reproduce those fine lattice results.

Note that $m_0$ is relatively small (close to zero) as a reflection
of $m_{\rm critical}$ for Wilson-type fermions lying close to
$m_{\rm charm}$ for our lattice spacing.  The significance of the
error in $m_0$ can be estimated from $J^{(3)}_{1,1}$ times the error
in $m_0$ from the average coarse data, giving a 4\% effect of the
error in $m_0$ on the resulting heavy-heavy, spin-averaged mass.

These better defined results for the case of the three-parameter
action demonstrate that the singularity in the matrix that must
be inverted to solve for these heavy quark parameters has disappeared.
For completeness we list the eigenvalues and eigenvectors of the
$3 \times 3$ matrix $(J^{(3)})^T J^{(3)}$ to be contrasted with the
singular results found for the four-parameter case in
Eqs.~\ref{eq:eigenvalues_4} and \ref{eq:eigenvectors_4}:
\begin{eqnarray}
\{  9.77(15), \; 1.41(10), \; 0.00026(4)\},
\label{eq:eigenvalues_3}
\end{eqnarray}
with corresponding eigenvectors
\begin{equation}
\begin{array}{rr@{.}lr@{.}lr@{.}ll}
\{ &0&824(4),  &-0&2157(11), & 0&524(6) &\},\nonumber\\
\{ &-0&504(8), & 0&142(7),   & 0&852(4) &\},         \\
\{ &0&258(4),  & 0&9661(11), &-0&008(7) &\}.\nonumber
\end{array}
\label{eq:eigenvectors_3}
\end{equation}
A comparison of Eqs.~\ref{eq:eigenvalues_4} and \ref{eq:eigenvectors_4} with
Eqs.~\ref{eq:eigenvalues_3} and \ref{eq:eigenvectors_3} shows that
the first two large eigenvalues and corresponding eigenvectors are
changed very little by the reduction from four to three parameters.

Next we would like to examine the contribution to systematic
error due to ignoring the quadratic terms in our analysis. Using
our 24 Cartesian data sets, we can calculate the both the first
($J$-matrix) and second derivatives (a quadratic matrix $Q$)
directly, without using a fitting procedure.   The resulting simple
Taylor expansions around the center point are:
\begin{equation}\label{eq:quadratic}
Y_q^n = Y^{(3),14}+ J^{(3)}\vdot (X^{(3),n} -
X^{(3),14}) + \frac{1}{2} (X^{(3),n} - X^{(3),14}) \vdot Q \vdot (X^{(3),n} - X^{(3),14})
\end{equation}
where $Q$ is the $3 \times 3$ tensor of second-derivatives and
$n$ runs from 43 to 66 (including only the Cartesian data sets). We can
now estimate how much our resulting parameters $X$ depend on the quadratic
terms and get a reasonable estimate of the systematic error.

Using this quadratic approximation, we determine the best-fit,
coarse lattice parameters $X^{(3)}_{\cal C}$ by minimizing
\begin{eqnarray}\label{eq:quadraticChi}
\chi_{{\cal F},\, q}^2&=&\Bigl(Y_{{\cal F}}-Y^{(3),14}
-J^{(3)}\vdot (X^{(3)}_{\cal C}-X^{(3),14}) \\
&& \hskip 1.0in    - 1/2 (X^{(3)}_{\cal C}-X^{(3),14})^T \vdot
                         Q^{(3)} \vdot (X^{(3)}_{\cal C}-X^{(3),14}) \Bigr)^T
\nonumber \\
&&\vdot W_{\cal F}^{-1} \vdot\Bigl(Y_{{\cal F}}-Y^{(3),14}-J^{(3)}\vdot
(X^{(3)}_{\cal C}-X^{(3),14}) \nonumber \\
&& \hskip 1.0in    - 1/2 (X^{(3)}_{\cal C}-X^{(3),14})^T \vdot Q^{(3)}
               \vdot (X^{(3)}_{\cal C}-X^{(3),14}) \Bigr). \nonumber
\end{eqnarray}
The result is $(X^{(3)}_{\cal C})^T = \{m_0,c_{\rm P},\zeta\}
=\{0.034(8),1.50(3),1.035(5)\}$, now including the effects of quadratic
terms.  Comparing these numbers with those in Eq.~\ref{eq:Xout_3} from
the linear approximation one sees that the quadratic contributions to
the results are buried in statistical noise.  Therefore, we will not
include contributions to the possible systematic errors coming the
neglect of these quadratic terms in the analysis.

The systematic errors enter as:  (a) We use $(ma)^2 = 0.2^2$ or 4\% as
an estimate of the heavy quark discretization errors from domain
wall fermion calculation on the fine lattice.  (b) The remaining RHQ
heavy quark discretization effect on the coarse lattice are
given by $(a\vec p)^2 = (\alpha_s(\mu = 1 GeV)ma)^2 \approx 0.004$.
(c) Finally we estimate 1.3\% as the systematic error arising from the
matching of the spatial volumes of fine and coarse lattices.  Therefore,
adding these three systematic errors in quadrature gives our final
coefficients:  $(X^{(3)}_{\cal C})^T=\{m_0, c_{\rm P}, \zeta\}
= \{0.037(26)(13), 1.50(9)(6),1.029(14)(40)\}$ where the first error
shown is statistical and the second systematic.

In our analysis, we have determined three parameters in the action
by requiring that seven physical quantities agree between the coarse
and fine lattices.   Can we match fewer physical quantities between
the coarse and fine lattice spacing calculations and obtain the same
result?  Table~\ref{tab:CoarseThreeParDataComp} summarizes the
results for various choices of the quantities being matched.  As we
can see, all the different choices give consistent values for our three
action parameters, agreeing within one $\sigma$.  Thus, we have very
consistent results for different choices of calculated quantities which
provides a numerical demonstration of the validity of the heavy quark
version of the Symanzik improvement program being implemented here.

Let us focus on two choices of measurements: index ``E'' using all
seven measurements and index ``B'' using only the results from
heavy-heavy data.  One might hope that the more measurements we
include in the analysis, the smaller the resulting errors will be.
However, it should be recognized that the cost in computer time of
making the additional measurements involving light quarks is high.
As we can see, despite its considerable added cost, the index ``E''
set makes only a small improvement on the statistical errors.  It
may be more sensible to double the number of configurations and
focus exclusively on the heavy-heavy system in future calculations.


\section{Comparisons with Other Approaches}
\label{Sec:comp}

In this section we compare the parameters $m_0$, $\zeta$ and
$c_P$ determined here for our $1/a=3.6$ GeV effective heavy
quark theory with the similar parameters determined by other
methods.  This serves both as an approximate check of the
results determined here and an opportunity to compare
perturbative and non-perturbative methods.

\subsection{Determining the bare mass $m_0$}\label{subsec:m0}

We first consider the bare mass $m_0$.  In most treatments
this parameter is related to a continuum, ``physical'' quark
mass by a combination of a shift coming from the intrinsic
chiral symmetry breaking of Wilson fermions and a multiplicative
renormalization factor $Z_m$:
\begin{equation}
m(\mu) = Z_m (m_0 - m_{\rm crit})/a
\label{eq:z_m}
\end{equation}
where $m_{\rm crit}$ locates the value of $m_0$ at which the pion
mass vanishes and $m(\mu)$ represents a continuum quark mass,
specified by a renormalization condition imposed at the energy
scale $\mu$.  In the discussion below we will use the
$\overline{\rm MS}$ scheme and $\mu=2.0$ GeV.  We have introduced
an explicit factor of the inverse lattice spacing in Eq.~\ref{eq:z_m}
to give the continuum quark mass its proper units.

We begin by determining the value of $m(\mu)$ which corresponds
to the $m_f=0.2$ input mass used in our reference, $\beta=6.638$ domain
wall fermion calculation.   For domain wall fermions Eq.~\ref{eq:z_m}
also applies but $m_0$ should be replaced by $m_f$ and $-m_{\rm crit}$
by $m_{\rm res}$, a measure of residual domain wall fermion chiral
symmetry breaking that is sufficiently small that it will be
neglected here.  While a quenched, $\beta=6.638$, domain wall
fermion value for $Z_m$ is not known, the value $Z_m \approx 1.59$
obtained at $\beta=6.0$~\cite{Blum:2001sr} may not be too far off.
(The results presented in Ref~\cite{Dawson:2002nr} can be used to
compare $Z_m$ evaluated with the DBW2 action at two very different
lattice spacings, $1/a=1.3$ GeV and $1/a=2.0$ where a change of less
than 3\% is seen.)  Thus, we will assume the calculations described
in this paper correspond to $\mu^{\overline{\rm MS}}
(\mu=2.0\mbox{GeV})=1.72$ GeV.  (This large value suggests
that our choice for $m_f$ on the fine lattice may be somewhat larger
than is appropriate for the charm quark mass.)

To relate this result to the value of $m_0$ expected in our
$\beta=6.351$ calculation we next determine $m_{\rm crit}$.
The critical quark mass can be estimated using both perturbative and
non-perturbative methods.  The two-loop, perturbative value for
$m_{\rm crit}$ for the Wilson gauge and clover fermion actions
has been obtained in Ref.~\cite{Panagopoulos:2001fn}:
\begin{eqnarray}\label{eq:PTKcritClover}
m_{\rm crit} & = & g^2 \Sigma^{(1)} + g^4 \Sigma^{(2)}\\
\Sigma^{(1)}& = &\frac{N_c^2-1}{N_c} \left( -0.1628571 \right. \nonumber\\
&+& \left. 0.04348303 c_{\rm SW} + 0.0180958 c_{\rm SW}^2  \right) \\
\Sigma^{(2)}& =& (N_c^2 -1)\left[\left(-0.017537 +
\frac{1}{N_c^2} 0.016567+
\frac{N_f}{N_c}0.00118618\right)\right.\nonumber\\
&+& \left( 0.002601 - \frac{1}{N_c^2} 0.0005597 -
\frac{N_f}{N_c}0.0005459\right) c_{\rm SW} \nonumber\\
&+& \left( - 0.0001556 +\frac{1}{N_c^2} 0.0026226 +
\frac{N_f}{N_c}
0.0013652\right) c_{\rm SW}^2 \nonumber\\
&+& \left(-0.00016315 + \frac{1}{N_c^2} 0.00015803-
\frac{N_f}{N_c}
0.00069225\right) c_{\rm SW}^3 \nonumber\\
&+& \left. \left(-0.000017219 +\frac{1}{N_c^2} 0.000042829 -
\frac{N_f}{N_c} 0.000198100\right) c_{\rm SW}^4\right],
\end{eqnarray}
where the number of fermion flavors $N_f  = 0$ in the quenched
approximation and the number of colors $N_c = 3$.  We can
obtain the coefficient, $c_{\rm SW}$, of the clover term
from the non-perturbative result of Ref.~\cite{Luscher:1996jn}:
\begin{eqnarray}\label{eq:NPcsw_1}
c_{\rm SW} = \frac{1 -0.656 g^2 - 0.152g^4 - 0.054 g^6}{1-0.922
g^2},
\end{eqnarray}
where $g$ is the lattice coupling constant and $\beta=6/g^2$.
This gives $c_{\rm SW} = 1.544$ on a $\beta = 6.351$ lattice
and $m_{\rm crit} = -0.219$.

An alternative way of computing $m_{\rm crit}$ is to take the
non-perturbative, the ALPHA collaboration measurement of
$\kappa_{\rm crit}$ (for example from Table~1 in
Ref.~\cite{Luscher:1996jn}) and parameterize it as a function of
coupling constant:
\begin{eqnarray}\label{eq:NPcsw_2}
\kappa_{\rm crit}&=&\frac{0.130287 - 0.239546 g^2 +
0.111829g^4}{1- 1.84915 g^2 + 0.868181 g^4}
\end{eqnarray}
for $ 6.0 \le \beta \le 7.4$ and use $m_{\rm crit} = \frac{1}{2
\kappa_{\rm crit} } -4$.  This gives $m_{\rm crit} = -0.317$.  We will
adopt this latter, non-perturbative value as being more accurate.

Finally, in order to invert Eq.~\ref{eq:z_m} to obtain the expected
value of $m_0$ which can be compared with our results we require the
appropriate factor $Z_m$ for our rather fine $\beta=6.351$ lattice
and relatively large, $m^{\overline{\rm MS}}(\mu=2.0\mbox{GeV})=1.72$ GeV
quark mass.  For this comparison, we can avoid the extra translation
to and from the $\overline{\rm MS}$ scheme by directly comparing
quantities calculated in the RI scheme at $\mu=2.0$ GeV.  From Tables I
and II of Ref.~\cite{Blum:2001sr} we determine $Z_m^{RI}({\rm DWF})=1.81$.
We will use a similar non-perturbative value
$Z_m^{RI}({\rm SW})=1/Z_S^{{\rm NPM}}=1.82$ extracted from Table 1 of
Ref.~\cite{Becirevic:1998yg}.  This value is only approximate for our
situation since it was obtained for light quark masses and on a coarser,
$\beta=6.2$, lattice.  Using these values we obtain
$m_0 = Z_m^{RI}({\rm DWF})/Z_m^{RI}({\rm SW}) \cdot m_f + m_{\rm crit} = -0.018$
in units of $1/a=3.6$ GeV.  Since the light quark value for
$Z_m^{RI}({\rm SW})$ chosen in this estimate will have $O(ma)$ errors, we
expect systematic errors of size $\sim 0.1$, and should view the agreement
between this $m_0=-0.018$ estimate and the $m_0=0.036(34)$ result in
Eq.~\ref{eq:Xout_3} as very reasonable.

\subsection{One-loop RHQ coefficients}
\label{subsec:oneLoopComp}

We now compare our non-perturbative result for the remaining
parameters $\zeta$ and $c_P$ with the one-loop perturbative
calculations carried out by M.~Nobes~\cite{Nobes:thesis} for
the closely related quantities, $\zeta$, $c_B$ and $c_E$
appearing in the Fermilab action.

These one-loop coefficients of Fermilab action were calculated
using automated perturbation theory techniques from the scattering
of a quark off of a background chromo-magnetic(electric)
field\cite{Nobes:thesis}. The calculations are done on the
lattice and in the continuum and the comparison used to determine
the lattice parameters.

The analytic tree-level coefficients (after being translated into
our notation for the action) are:
\begin{eqnarray}\label{eq:treeFermi}
\zeta^{[0]} &=& \sqrt{\left(\frac{m_0 (2+m_0)}{4 (1+m_0)}\right)^2
+\frac{m_0 (2+m_0)}{2 \ln(1+m_0)}}-\frac{ m_0 (2+m_0)}{4 (1+m_0)}\nonumber \\
c_B^{[0]} &=&  \zeta^{[0]} \nonumber\\
c_E^{[0]} &=& \zeta^{[0]} \left(\frac{(\zeta^{[0]})^2-1}{m_0
(2+m_0)}+\frac{\zeta^{[0]}}{(1+m_0)}+\frac{ m_0
(2+m_0)}{4(1+m_0)^2}\right).
\end{eqnarray}

Next, the one-loop result for $\zeta^{[1]}$, is given by the
formula:
\begin{eqnarray}\label{eq:oneloopZeta}
\zeta^{[1]} &=& - \left(1+g_0^2
Z_{M_2}^{[1]}\right)\frac{(\zeta^{[0]})^2+\zeta^{[0]}\sinh\left(\ln(1+m_0)\right)}
         {\zeta^{[0]}+\sinh\left(\ln(1+m_0)\right)}.
\end{eqnarray}
We use this formula and the numerical one-loop results from
Tables~6.1 and 6.2 in Ref.~\cite{Nobes:thesis} and perform an
error-weighted fit to the three functions of interest, $\zeta(m_0a)$,
$c_B(m_0a)$ and $c_E(m_0a)$ with expressions of the form
\begin{eqnarray}\label{eq:parmFermiCoeff}
X^{[1]}=\frac{\sum_{i=0}^{3} a_i m_0^i}{1+\sum_{i=1}^{3} d_i m_0^i}
\end{eqnarray}
where $X$ represents $c_B$, $c_E$ and $\zeta$ while the $a$'s and
$d$'s are listed in Table~\ref{tab:FermiOneLoop}.
This fit implies that at $m_0=0.036$, the coefficients are
$c_B=1.261$, $c_E=1.246$ and $\zeta=1.003$, $\approx 1.4\sigma$
lower than our non-perturbatively determined coefficients:
$(X^{(3)}_{\cal C})^T=\{0.037(26)(13), 1.50(9)(6),1.029(14)(40)\}$.
(Since the results of Nobes have $c_B \approx c_E$ we can directly
compare the coefficients in his 4-parameter and our 3-parameter
lattice action.)

To see directly the effects of the differences between these
perturbative and non-perturbative coefficients, we should
compare the resulting spectra.  Although we did not use these
one-loop numbers in a spectrum calculation, we can use our
linear description of the dependence of the spectra on the
action parameters (the coefficients $J$ and $A$ of
Eq.~\ref{eq:FitCoarse}) to get a good idea of what the
resulting masses would be were we to use these one-loop
coefficients. We summarize the results in Table~\ref{tab:PredictedData}.
These are reasonably close to our non-perturbative results
with the largest discrepancy being the two hyperfine splittings
which are 25\% smaller when determined from the one-loop
coefficients.

There is a second, extensive perturbative calculation of the
one-loop, tadpole improved RHQ lattice action by the Tsukuba
group~\cite{Aoki:2003dg}.  However, because the Tsukuba
action uses five parameters with $r_s \ne \zeta$, we cannot make
a direct comparison.  While continuum field transformations
can be employed on the continuum effective Lagrangian to
prove that these 5-parameter and 3-parameters descriptions
should lead to the same continuum physics up to discrepancies
of order $(\vec p a)^2$, these transformations are not defined
for the lattice variables and cannot be used to relate the
one-loop coefficients of the Tsukuba action given in
Ref.~\cite{Aoki:2003dg} and those determined here.

\section{Summary and Outlook}
\label{SecFuture}

In this work, we have demonstrated that it is practical to
determine the coefficients of the relativistic heavy quark action
non-perturbatively through a finite-volume, step-scaling technique.
This has been done by matching various heavy-heavy and heavy-light
mass spectrum calculations on two quenched lattices.  The domain
wall fermion action is used on fine lattice, where $ma$ is relatively
small, while for the coarse lattice an improved relativistic heavy
quark action is used.   By comparing the finite-volume predictions
of these two theories we can then determine the coefficients of the
heavy quark action.  In order to simplify the analysis, we assumed
a linear relation between the parameters appearing in the heavy
quark action and the resulting mass spectrum.  The coefficients in
this linear relation were determined by computing the coarse-lattice
mass spectrum for a number of choices for the RHQ action.  We could
then use this linear relation to precisely determine those heavy
quark parameters which would yield the masses implied by the fine
lattice calculations.

We initially applied this matching technique to the four-parameter
version of the heavy quark action originally proposed in
Ref.~\cite{El-Khadra:1997mp}.  However, for this case, the
system of linear equations that must be solved was singular
within statistical errors and the resulting parameters, especially
the coefficients $c_B$ and $c_E$ very poorly determined.  This
lead us to search for possible redundancy in the four-parameter
action and recognize, as is discussed in detail in a companion
paper~\cite{3parAction} that a further field transformation was available
that could be used to set $c_E=c_B$, reducing the number of
independent parameters to three.  With this restriction the
problem of determining the relativistic heavy quark action is
well-posed and the coefficients can be accurately determined.
Our result for the bare mass, clover term and asymmetry between the
space and time derivatives is $\{m_0, c_{\rm P}, \zeta\}=\{0.037(26)(13),
1.50(9)(6),1.029(14)(40)\}$, where the first error is statistical and
the second systematic, excluding those coming from the quenched
approximation.  Finally, we included a quadratic term in the dependence
of our measured masses on the action parameters and obtained a result
consistent with the linear expansion.

We can easily decrease the statistical error by increasing the
number of configurations (here 100 were used) and reduce the
systematic error by starting with a finer lattice for the domain
wall fermion calculation.  Our use of the quenched calculation is
intended to provide a computationally inexpensive study of the
matching procedure.  The next step is a determination of the
coefficients in this relativistic heavy quark action, appropriate
for charm physics in full QCD.  As discussed in Sec.~\ref{sec:Strategy},
we can perform the same finite-volume, step-scaling procedure on
2+1 flavor dynamical lattices.  Since the long- and short-distance
physics can be treated separately, we can substantially reduce the
computational cost of such full QCD step scaling by using heavier
light quark sea masses in the earlier stage of matching, as long as
$m_{sea}$/$\Lambda_{QCD}$ are equal for each pair of systems being
matched.  Such a calculation should be practical on presently
available computers.

\section*{ACKNOWLEDGMENTS}
The authors would thank Peter Boyle, Paul Mackenzie, Sinya Aoki
and Yoshinobu Kuramashi for helpful discussions, Taku Izubuchi and
Koichi Hashimoto for their static quark potential code, Tanmoy
Bhattacharya for discussion of a similar, on-shell, finite volume
approach which he had considered earlier and members of the RBC
collaboration for their help throughout the course of this work.  In
addition, we thank Peter Boyle, Dong Chen, Mike Clark, Saul Cohen,
Calin Cristian, Zhihua Dong, Alan Gara, Andrew Jackson, Balint Joo,
Chulwoo Jung, Richard Kenway, Changhoan Kim, Ludmila Levkova,
Xiaodong Liao, Guofeng Liu, Robert Mawhinney, Shigemi Ohta, Konstantin
Petrov, Tilo Wettig and Azusa Yamaguchi for developing with us the
QCDOC machine and its software. This development and the resulting
computer equipment used in this calculation were funded by the
U.S. DOE grant DE-FG02-92ER40699, PPARC JIF grant PPA/J/S/1998/00756
and by RIKEN. This work was supported by DOE grant DE-FG02-92ER40699
and we thank RIKEN, Brookhaven National Laboratory and the U.S.
Department of Energy for providing the facilities essential for the
completion of this work.
\vspace{-0.1in}


\bibliography{06np_rhq}

\clearpage
\begin{table}[hbt]
\caption{\label{tab:RHQComp} Comparison of the conventions and/or
values used to specify the six terms in the improved lattice action
of Eq.~\ref{eq:RHQ}.  The top row identifies terms that appear in
Eq.~\ref{eq:RHQ}.  The next row lists our choice for the coefficient
of each term and the next two rows specify that same coefficient
written in the notation of the Fermilab~\cite{El-Khadra:1997mp} and
Tsukuba~\cite{Aoki:2001ra} papers.}
\begin{center}
\begin{tabular}{ccccccc}
\hline \hline
Action     & $\gamma^0 D^0$ \quad
               & $\vec \gamma \vec D$   \quad
                         & $-D_0^2$  \quad
                             & $-(\vec D)^2$  \quad
                                           & $\frac{i}{2}\sigma_{ij}F_{ij}$  \quad
                                                         &$i\sigma_{0i}F_{0i}$   \quad\\
\hline
This paper & 1 & $\zeta$ & 1 & $r_s$       & $c_B$       &$c_E$ \\
Fermilab   & 1 & $\zeta$ & 1 & $r_s \zeta$ & $c_B \zeta$ & $c_E \zeta$\\
Tsukuba    & 1 & $\nu$   & 1 & $r_s$       & $c_B$       & $c_E$ \\
\hline\hline
\end{tabular}
\end{center}
\end{table}

\begin{table}[hgt]
\caption{\label{tab:parametersA}  The specific choice of parameters
for the two sets of lattice configurations analyzed in this paper.}
\begin{center}
\begin{tabular}{l c c}
     & I                           & II               \\
\hline \hline
Volume & $24^3 \times 48$          & $16^3 \times 32$ \\
$1/a$    & 5.4 GeV                 & 3.6 GeV          \\
$L$      & 0.9 fm                  & 0.9 fm           \\
$\beta$  & 6.638                   & 6.351            \\
$ma$     & 0.2                     & 0.3              \\
Action   & DWF                     & RHQ
\end{tabular}
\end{center}
\end{table}

\begin{table}[hgt]
\caption{\label{tab:parametersB}  The choice of parameters for the
two sets of lattice configurations needed for the next step in
this step-scaling program.  Calculations with these parameters
are now underway but are not described in this paper.}
\begin{center}
\begin{tabular}{l c c}
     & III                         & IV               \\
\hline \hline
Volume & $24^3 \times 48$          & $16^3 \times 32$ \\
$1/a$    & 3.6 GeV                 & 2.4 GeV          \\
$L$      & 1.33 fm                 & 1.33 fm          \\
$\beta$  & 6.351                   & 6.074                \\
$ma$     & 0.3                     & 0.45             \\
Action   & RHQ                     & RHQ
\end{tabular}
\end{center}
\end{table}

\begin{table}
\caption{ Meson states created by local operators of the form
$\bar{\psi}\Gamma\psi$, labelled in spectroscopic notation. }
\label{tab:meson_op}
\begin{center}
\begin{tabular}{ccc}
\hline\hline
          $\Gamma$              & $^{2S+1}L_J$   & $J^{PC}$  \\
          \hline
          $\gamma_5$            & $^1S_0$        & $0^{-+}$  \\
          $\gamma_i$            & $^3S_1$        & $1^{--}$  \\
          $1$                   & $^3P_0$        & $0^{++}$  \\
          $\gamma_5\gamma_i$    & $^3P_1$        & $1^{++}$  \\
\hline\hline
\end{tabular}
\end{center}
\end{table}

\begin{table}[hbt]
\caption{\label{tab:fixedPar} Common parameters for each of the coarse
and fine data sets.  For both data sets $L=0.9$~fm while for the domain
wall fermion action we use $L_s = 12$ and $M_5 = 1.5$.  Here ``TBD'' indicates
a value to be determined in the matching proceedure being developed here.}
\begin{center}
\begin{tabular}{cccccccc}
\hline \hline %
Label & $\beta$ &       V          &  $S_{L}$ & $am_L$ & $S_H$ & $am_H$   &  $a^{-1}$(SQ pot.)\\
\hline %
{Fine}   & 6.638   &  $24^3\times48$  &  DWF     & 0.02  &  DWF  & 0.2       &    5.4~GeV \\
{Coarse} & 6.351 & $16^3\times32$   &  DWF    & 0.03 & RHQ  & TBD  &    3.6~GeV\\
\hline \hline %
\end{tabular}
\end{center}
\end{table}

\begin{table}[hbt]
\caption{\label{tab:FineData} Mass spectrum measured on the fine
lattice in units of $a^{-1}$
}
\begin{center}
\begin{tabular}{ccccccc}
\hline \hline %
              &  $m_{\rm PS}$ &  $m_{\rm V}$  &  $m_1/m_2$  &  $m_{\rm AV}$  & $m_{\rm S}$  &  
               \\
\hline %
light-light   &  0.175(3) &  0.233(5) &   --   &   --   &   --  \\
heavy-light   &  0.467(2) &  0.485(3) &   --   &   --   &   --  \\
heavy-heavy   &  0.716(1)&   0.728(1) &  1.02(2)&   0.810(5) & 0.799(4) \\
\hline \hline %
\end{tabular}
\end{center}
\end{table}

\begin{table}[hbt]
\caption{ Light-light hadron spectrum measured on the coarse lattice
in units of $a^{-1}$ and units of $3/2a$ to compare with the fine lattice
results in Table~\ref{tab:FineData}.}
\label{tab:CoarseDataLight}
\begin{center}
\begin{tabular}{cccc}
\hline\hline %
              & units & $m_{\rm PS}$ &  $m_{\rm V}$   \\
\hline %
light-light   & 1/a   & 0.259(6)     &  0.328(10)  \\
light-light   & 3/2a  & 0.173(4)     &  0.219(7)   \\
\hline \hline %
\end{tabular}
\end{center}
\end{table}

\begin{table}[hbt]
\caption{\label{tab:AllCoarsePara} Parameters used on the coarse lattice.} 
{\scriptsize
\begin{center}
\begin{minipage}{7cm}
\begin{tabular}{ccccc}
\hline \hline %
    Set \#   &   $m_0$       &    $c_B$   &     $c_E$  &    $\zeta$  \\
\hline %
 1 & 0. & 1.55206 & 1.45769 & 1.01281\\
 2 & 0.07 & 1.5474 & 1.42445 & 1.00063\\
 3 & 0.0426 & 1.55034 & 1.43843 & 1.00674\\
 4 & 0.0426 & 1.55034 & 1.43843 & 1.1\\
 5 & 0.0426 & 1.55034 & 1.43843 & 0.9\\
 6 & 0.0330029 & 1.60921 & 1.53843 & 1.04395\\
 7 & 0.0230029 & 1.60921 & 1.43843 & 1.04395\\
 8 & 0.0230029 & 1.60921 & 1.53843 & 1.04395\\
 9 & 0.0426 & 1.60921 & 1.43843 & 1.04395\\
 10 & 0.0426 & 1.55034 & 1.43843 & 1.00674\\
 11 & 0. & 1.55206 & 1.43843 & 1.01281\\
 12 & 0.0327893 & 1.51081 & 1.43843 & 1.03563\\
 13 & 0.0230029 & 1.51081 & 1.43843 & 1.03563\\
 14 & 0.0327893 & 1.51081 & 1.53843 & 1.03563\\
 15 & 0.01 & 1.70012 & 1.57429 & 1.02152\\
 16 & 0.003705 & 1.70862 & 1.5768 & 1.02349\\
 17 & 0.0138 & 1.71495 & 1.57871 & 1.02499\\
 18 & 0.02 & 1.71886 & 1.57991 & 1.02593\\
 19 & 0.03 & 1.72523 & 1.58188 & 1.02749\\
 20 & 0.08 & 1.70683 & 1.57627 & 1.02307\\
 21 & 0.09 & 1.71308 & 1.57815 & 1.0245\\
 22 & 0.1 & 1.71939 & 1.58007 & 1.02606\\
 23 & 0.101197 & 1.5031 & 1.93237 & 1.00317\\
 24 & 0.159393 & 1.45619 & 2.42673 & 0.999347\\
 25 & 0.21759 & 1.40929 & 2.9211 & 0.99552\\
 26 & 0.275786 & 1.36238 & 3.41547 & 0.991694\\
 27 & 0.333983 & 1.31548 & 3.90983 & 0.987867\\
 28 & 0.392179 & 1.26858 & 4.4042 & 0.984041\\
 29 & -0.132144 & 1.96391 & 0.619831 & 1.04758\\
 30 & -0.0642197 & 1.83627 & 1.09832 & 1.03553\\
 31 & -0.0302573 & 1.77244 & 1.33756 & 1.02951\\
 32 & -0.00987993 & 1.73415 & 1.4811 & 1.0259\\
 33 & 0.0172899 & 1.68309 & 1.6725 & 1.02108\\
\hline \hline %
\end{tabular}
\end{minipage}
\begin{minipage}{6cm}
\begin{tabular}{ccccc}
\hline \hline %
    Set \#   &   $m_0$       &    $c_B$   &     $c_E$  &    $\zeta$  \\
\hline %
 34 & 0.0376673 & 1.6448 & 1.81604 & 1.01747\\
 35 & 0.043216 & 1.5843 & 0.635194 & 1.04415\\
 36 & 0.0279157 & 1.62166 & 1.49005 & 1.04385\\
 37 & 0.038103 & 1.59676 & 1.58681 & 1.04405\\
 38 & 0.0228413 & 1.63412 & 1.44167 & 1.04375\\
 39 & 0.0124458 & 1.7295 & 1.23541 & 1.04295\\
 40 & 0.0124458 & 1.7295 & 1.61502 & 1.04295\\
 41 & 0.0124458 & 1.7295 & 0.855798 & 1.04295\\
 42 & 0.027386 & 1.70923 & 1.30737 & 1.03874\\
 43 & 0.132789 & 1.61081 & 1.83843 & 1.05563\\
 44 & -0.0672107 & 1.61081 & 1.83843 & 1.05563\\
 45 & 0.132789 & 1.41081 & 1.83843 & 1.05563\\
 46 & -0.0672107 & 1.41081 & 1.83843 & 1.05563\\
 47 & 0.132789 & 1.61081 & 1.23843 & 1.05563\\
 48 & -0.0672107 & 1.61081 & 1.23843 & 1.05563\\
 49 & 0.132789 & 1.41081 & 1.23843 & 1.05563\\
 50 & -0.0672107 & 1.41081 & 1.23843 & 1.05563\\
 51 & 0.132789 & 1.61081 & 1.83843 & 1.01563\\
 52 & -0.0672107 & 1.61081 & 1.83843 & 1.01563\\
 53 & 0.132789 & 1.41081 & 1.83843 & 1.01563\\
 54 & -0.0672107 & 1.41081 & 1.83843 & 1.01563\\
 55 & 0.132789 & 1.61081 & 1.23843 & 1.01563\\
 56 & -0.0672107 & 1.61081 & 1.23843 & 1.01563\\
 57 & 0.132789 & 1.41081 & 1.23843 & 1.01563\\
 58 & -0.0672107 & 1.41081 & 1.23843 & 1.01563\\
 59 & 0.132789 & 1.51081 & 1.53843 & 1.03563\\
 60 & -0.0672107 & 1.51081 & 1.53843 & 1.03563\\
 61 & 0.0327893 & 1.61081 & 1.53843 & 1.03563\\
 62 & 0.0327893 & 1.41081 & 1.53843 & 1.03563\\
 63 & 0.0327893 & 1.51081 & 1.83843 & 1.03563\\
 64 & 0.0327893 & 1.51081 & 1.23843 & 1.03563\\
 65 & 0.0327893 & 1.51081 & 1.53843 & 1.05563\\
 66 & 0.0327893 & 1.51081 & 1.53843 & 1.01563\\
\hline \hline %
\end{tabular}
\end{minipage}
\end{center}
}
\end{table}

\begin{table}[hbt]
\caption{\label{tab:AllCoarseData} Mass spectrum measured on the coarse
lattice in units of $a^{-1}$ (where ``sa'' is spin averaged,
``hs'' is hyperfine splitting, ``soa'' is spin-orbit averaged,
``sos'' is spin-orbit splitting; ``hl'' is heavy-light and ``hh''
is heavy-heavy).}
\begin{center} {\scriptsize
\begin{tabular}{cccccccc}
\hline \hline %
  Set \# & $m^{hh}_{\rm sa}$ & $m^{hh}_{\rm hs}$ & $m^{hl}_{\rm sa}$ & $m^{hl}_{\rm hs}$ &
  $m^{hh}_{\rm sos}$ & $m^{hh}_{\rm soa}$& $m_1/m_2$ \\
\hline%
1 &1.0137(17) & 0.0174(6) & 0.679(5) & 0.0254(17) &0.020(5) &    1.135(14) &0.988(21)\\
2 &1.1314(15) & 0.0152(5) & 0.742(4) & 0.0221(15) & 0.017(4) &    1.251(13) & 0.949(18)\\
3 &1.0872(16) & 0.0160(5) & 0.718(5) & 0.0233(15) &0.018(4) &    1.207(14) & 0.966(19)\\
4 &1.1755(16) & 0.0153(5) & 0.762(5) &0.0228(15) & 0.017(4) &    1.298(13) & 1.075(21)\\
5 &0.9776(16) & 0.0169(5) &0.664(5) & 0.0241(16) & 0.020(5) &    1.094(14) & 0.842(16)\\
6 &1.0724(17) &0.0171(5) & 0.709(5) & 0.0248(16) & 0.019(5) &    1.193(14) &1.021(21)\\
7 &1.0840(15) & 0.0170(5) & 0.716(4) &0.0246(16) & 0.018(4) &     1.208(12) & 1.009(17)\\
8 & 1.0622(15) & 0.0176(5) & 0.704(4) & 0.0254(16) &0.025(4) &    1.190(12) & 1.016(18)\\
9 &1.1177(15) & 0.0165(5) & 0.734(4) &0.0238(15) & 0.017(4) &    1.241(12) & 1.002(16)\\
10 &1.0949(14) & 0.0161(5) &0.723(4) & 0.0234(15) & 0.017(4) &    1.217(12) & 0.956(15)\\
11 &1.0255(15) &0.0174(5) & 0.686(4) & 0.0254(16) & 0.019(4) &    1.149(12) &0.980(17)\\
12 &1.1148(15) & 0.0157(5) & 0.733(4) &0.0232(15) & 0.017(4) &     1.238(12) & 0.992(16)\\
13 &1.0981(15) & 0.0160(5) & 0.724(4) & 0.0236(15) &0.017(4) &    1.221(12) & 0.995(16)\\
14 &1.0937(15) & 0.0162(5) & 0.721(4) &0.0239(15) & 0.018(4) &    1.216(12) & 0.998(16)\\
15 &0.9384(19) & 0.0204(7) &0.638(5) & 0.0287(19) & 0.024(6) &    1.060(15) & 1.016(23)\\
16 &0.9637(18) &0.0199(6) & 0.652(5) & 0.0280(18) & 0.023(5) &    1.085(15) &1.013(23)\\
17 &0.9822(18) & 0.0195(6) & 0.661(5) &0.0275(18) & 0.022(5) &    1.103(15) & 1.012(22)\\
18 &0.9935(18) &0.0193(6) & 0.667(5) &0.0273(18) &0.022(5) &    1.114(14) & 1.011(22)\\
19 &1.0115(18) & 0.0190(6) & 0.677(5) &0.0268(17) & 0.021(5) &    1.132(14) & 1.009(22)\\
20 &1.1019(16) & 0.0171(5) &0.725(5) & 0.0242(16) & 0.019(5) &    1.221(13) & 0.986(20)\\
21 &1.1187(16) &0.0169(5) & 0.734(5) & 0.0239(16) & 0.018(4) &    1.237(13) &0.985(19)\\
22 &1.1353(16) & 0.0167(5) & 0.743(5) &0.0235(15) & 0.018(4) &     1.254(13) & 0.984(19)\\
23 &1.0809(16) & 0.0164(5) & 0.714(5) & 0.0240(16) &0.019(5) &    1.196(14) & 0.973(19)\\
24 &1.0521(16) & 0.0175(6) & 0.699(5) &0.0255(16) & 0.020(5) &    1.164(14) & 0.988(20)\\
25 &0.9985(17) & 0.0195(6) &0.670(5) & 0.0279(18) & 0.023(6) &    1.107(14) & 1.013(22)\\
26 &0.9143(19) &0.0229(8) & 0.624(5) & 0.0318(20) & 0.027(7) &    1.021(15) &1.054(27)\\
27 &0.7896(23) & 0.0295(11) & 0.557(5) &0.0384(25) & 0.035(9) &      0.895(18) & 1.12(4)\\
28 &0.606(3) & 0.0469(28) & 0.460(6) & 0.052(4) &0.053(18) &    0.708(24) & 1.21(10)\\
29 &0.8593(22) & 0.0235(8) & 0.596(5) &0.0316(21) & 0.026(7) &    0.993(17) & 1.049(28)\\
30 &0.9229(20) & 0.0212(7) &0.630(5) & 0.0293(19) & 0.024(6) &    1.050(15) & 1.027(25)\\
31 &0.9461(19) &0.0204(7) & 0.642(5) & 0.0286(19) & 0.023(6) &    1.070(15) &1.019(24)\\
32 &0.9573(19) & 0.0201(7) & 0.648(5) &0.0282(18) & 0.023(6) &     1.079(15) & 1.016(23)\\
33 &0.9693(18) & 0.0197(6) & 0.655(5) & 0.0279(18) &0.023(5) &    1.089(15) & 1.011(23)\\
\hline \hline %
\end{tabular}
}
\end{center}
\end{table}
\clearpage
\begin{center} {\scriptsize
\begin{tabular}{cccccccc}
\hline \hline %
  Set \# & $m^{hh}_{\rm sa}$ & $m^{hh}_{\rm hs}$ & $m^{hl}_{\rm sa}$ & $m^{hl}_{\rm hs}$ &
 $m^{hh}_{\rm sos}$ & $m^{hh}_{\rm soa}$ & $m_1/m_2$ \\
\hline%
34 &0.9760(18) & 0.0195(6) & 0.658(5) &0.0277(18) & 0.023(5) &    1.095(14) & 1.009(22)\\
35 &1.2455(15) & 0.0136(4) &0.802(4) & 0.0195(13) & 0.013(4) &    1.372(13) & 0.976(18)\\
36 &1.0711(17) &0.0171(5) & 0.708(5) & 0.0248(16) & 0.019(5) &    1.192(14) &1.020(21)\\
37 &1.0736(17) & 0.0170(5) & 0.709(5) &0.0248(16) & 0.019(5) &     1.194(14) & 1.022(21)\\
38 &1.0696(17) & 0.0172(5) & 0.707(5) & 0.0248(16) &0.019(5) &    1.191(14) & 1.019(21)\\
39 &1.0700(17) & 0.0176(5) & 0.708(5) &0.0249(16) & 0.019(5) &    1.194(14) & 1.013(21)\\
40 &0.9864(18) & 0.0198(6) &0.663(5) & 0.0279(18) & 0.022(5) &    1.108(15) & 1.036(23)\\
41 &1.1359(16) &0.0162(5) & 0.743(5) & 0.0228(15) & 0.017(4) &    1.262(13) &0.997(20)\\
42 &1.0824(17) & 0.0173(5) & 0.715(5) &0.0245(16) & 0.019(5) &     1.205(14) & 1.007(21)\\
43 &1.1866(15) & 0.0157(5) & 0.769(4) & 0.0227(15) &0.017(4) &    1.303(13) & 1.019(20)\\
44 &0.8149(23) & 0.0245(9) & 0.571(5) &0.0340(23) & 0.030(7) &    0.937(17) & 1.10(3)\\
45 &1.2271(15) & 0.0138(4) &0.790(4) & 0.0209(14) & 0.015(4) &    1.344(13) & 1.012(19)\\
46 &0.8705(22) &0.0212(8) & 0.601(5) & 0.0309(21) & 0.027(6) &    0.993(16) &1.090(28)\\
47 &1.3000(14) & 0.0136(4) & 0.830(4) &0.0195(13) & 0.014(4) &     1.420(12) & 0.986(18)\\
48 &0.9691(19) & 0.0189(6) & 0.654(5) & 0.0274(18) &0.022(5) &    1.095(15) & 1.052(24)\\
49 &1.3356(14) & 0.0119(3) & 0.849(4) &0.0179(12) & 0.012(3) &    1.456(12) & 0.979(17)\\
50 &1.0164(18) & 0.0165(5) &0.679(5) & 0.0250(17) & 0.020(5) &    1.143(14) & 1.045(22)\\
51 &1.1473(15) &0.0160(5) & 0.749(4) & 0.0230(15) & 0.018(4) &    1.263(13) &0.973(19)\\
52 &0.7632(24) & 0.0257(9) & 0.545(5) &0.0354(24) & 0.032(8) &     0.884(17) & 1.06(3)\\
53 &1.1894(15) & 0.0141(4) & 0.772(4) & 0.0210(14) &0.016(4) &    1.305(13) & 0.967(18)\\
54 &0.8218(22) & 0.0221(8) & 0.576(5) &0.0319(21) & 0.028(7) &    0.943(16) & 1.045(28)\\
55 &1.2662(14) & 0.0138(4) &0.813(4) & 0.0196(13) & 0.014(4) &    1.385(13) & 0.940(17)\\
56 &0.9272(19) &0.0194(6) & 0.633(5) & 0.0279(18) & 0.023(6) &    1.052(15) &1.004(23)\\
57 &1.3030(14) & 0.0120(3) & 0.833(4) &0.0179(12) & 0.013(3) &     1.422(12) & 0.934(17)\\
58 &0.9764(18) & 0.0169(6) & 0.659(5) & 0.0254(17) &0.020(5) &    1.102(15) & 0.997(21)\\
59 &1.2495(15) & 0.0137(4) & 0.803(4) &0.0195(13) & 0.014(4) &    1.367(12) & 0.978(18)\\
60 &0.9026(20) & 0.0200(7) &0.619(5) & 0.0284(19) & 0.023(6) &    1.027(15) & 1.052(24)\\
61 &1.0626(17) &0.0172(5) & 0.704(4) & 0.0242(16) & 0.018(5) &    1.184(14) &1.016(20)\\
62 &1.1071(17) & 0.0151(5) & 0.728(4) &0.0221(15) & 0.016(4) &     1.228(13) & 1.009(19)\\
63 &1.0140(18) & 0.0179(6) & 0.678(4) & 0.0256(17) &0.020(5) &    1.134(14) & 1.034(21)\\
64 &1.1453(16) & 0.0148(5) & 0.748(4) &0.0212(14) & 0.015(4) &    1.267(13) & 0.995(19)\\
65 &1.1051(17) & 0.0160(5) &0.726(4) & 0.0230(15) & 0.017(4) &    1.226(13) & 1.036(20)\\
66 & 1.0656(17) &0.0163(5) & 0.706(4) & 0.0233(15) & 0.017(4) &
1.186(13) &0.989(19)\\
\hline \hline %
\end{tabular}
}
\end{center}
\clearpage

\begin{table}[hbt]
\caption{\label{tab:CoarseThreeParDataComp} The resulting coarse-lattice
paramters obtained by matching various combinations of physical
quantities between the coarse and fine lattices.  Different choices of 
for the quanties to be matched give action parameters consistent with 
each other within one $\sigma$, showing the consistency of this heavy 
quark improvement program.}
\begin{center}
\begin{tabular}{ccc}
\hline \hline %
Measurement Index  &   Data used  &    $X_{\cal C}^{(3)}$ \\
\hline
 A  &  $m^{hh}_{\rm sa}$, $m^{hh}_{\rm hs}$, $m^{hh}_{\rm sos}$, $m_1/m_2$                     &
             \{ 0.07(4), 1.67(13), 1.030(14)\} \\
 B  &  ``A'' $+  m^{hh}_{\rm soa}$                                                             &
             \{ 0.04(3), 1.56(10), 1.034(12)\} \\
 C  &  $m^{hh}_{\rm sa}$, $m^{hh}_{\rm hs}$, $m^{hl}_{\rm sa}$,  $m^{hl}_{\rm hs}$, $m_1/m_2$  &
             \{ 0.06(4), 1.62(13), 1.032(14)\} \\
 D  &  ``D'' $+  m^{hh}_{\rm sos}$                                                             &
             \{ 0.04(3), 1.56(10), 1.034(12)\} \\
 E  &  all                                                                                     &
             \{ 0.03(3), 1.53(10), 1.035(12)\} \\
\hline\hline
\end{tabular}
\end{center}
\end{table}

\begin{table}[htb]
\caption{\label{tab:FermiOneLoop} Parametrization of the one-loop
coefficients of the Fermilab action using Eq.~\ref{eq:parmFermiCoeff}.}
\begin{center}
\begin{tabular}{cccccccc}
\hline \hline %
     & $a_0$ & $a_1$ & $a_2$ & $a_3$ & $d_1$&$d_2$ & $d_3$ \\
\hline %
$\zeta^{[1]}$ &0.00029923 & 0.00124977& 0.163759 &0.0258287& 5.10243  & 1.65713 & 0.00633212 \\
$c_E^{[1]}$ & 0.270419 & 0.431474 & 0.162718 &  0.00212438 & 1.87436 & 0.319194  & 0.00619183\\
$c_B^{[1]}$ & 0.271519 & 0.0122322 & $-0.000039117$ &  0 &  0.0565955 & 0 & 0\\
\hline\hline
\end{tabular}
\end{center}
\end{table}

\begin{table}[hbt]
\caption{\label{tab:PredictedData} The expected coarse lattice results for various 
choices of coefficients, $X_{\cal C}$, in the heavy quark effective action.   Here 
the linear approximation of Eq.~\ref{eq:FitCoarse}, with coefficients $A$ and $J$ 
determined from data sets \#43-66, is being used to predict the corresponding 
physical masses.}
\begin{center}
\begin{tabular}{ccccccccc}
\hline \hline %
Parameter & $m^{hh}_{\rm sa}$ & $m^{hh}_{\rm hs}$ & $m^{hl}_{\rm
sa}$ & $m^{hl}_{\rm hs}$ &
 $m^{hh}_{\rm sos}$ & $m^{hh}_{\rm soa}$ & $m_1/m_2$ \\
\hline%
$X_{\cal C}^{(3)}$ & 1.0854(16)& 0.0165(5) & 0.716(4) & 0.0239(16) & 0.019(4) & 1.206(13) & 1.000(20)\\
$X_{\cal C}^{Fermilab}$ &  1.1700(15) & 0.0124(4) & 0.763(4) & 0.0194(13) & 0.014(4) & 1.292(13) & 0.943(16) \\
scaled $Y_{\cal F}$ & 1.0881(16) & 0.0175(6) & 0.725(3) & 0.0267(13) &  0.017(3) &  1.211(8) & 1.002(23)\\
\hline\hline
\end{tabular}
\end{center}
\end{table}

\clearpage

 \vskip -0mm

\end{document}